\documentclass[
 article,
 twocolumn,
 groupedaddress,
 showpacs,
 preprintnumbers,
 amsmath,
 amssymb,
 aps,
 prstab,
 floatfix,
]{revtex4-1}

\usepackage{graphicx}	
\usepackage{hyperref}
\usepackage{verbatim}					
\usepackage{dcolumn}                    
\usepackage[ruled,vlined]{algorithm2e}
\include{pythonlisting}
\usepackage{xcolor}

\newcommand{\pd}{\partial}				    

\newcommand{\const}{\mathrm{const}}         
\newcommand{\K}{\mathcal{K}}				
\newcommand{\m}{\mathcal{M}}				

\newcommand{\ds}{\displaystyle}				

\begin{document}

\title{Integrable symplectic maps with a polygon tessellation}
\author{T.~Zolkin}
\email{zolkin@fnal.gov}
\affiliation{Fermilab, PO Box 500, Batavia, IL 60510-5011}
\author{Y.~Kharkov}
\affiliation{Joint Center for Quantum Information and Computer Science,
NIST/University of Maryland, College Park, MD 20742}
\affiliation{Joint Quantum Institute,
NIST/University of Maryland, College Park, MD 20742}
\thanks{Currently at AWS Quantum Technologies}
\author{S.~Nagaitsev}
\email{nsergei@jlab.org}
\affiliation{Jefferson Lab, Newport News, VA 23606}
\affiliation{Old Dominion University, Norfolk, VA 23529}
\date{\today}

\begin{abstract}

Identifying integrable dynamics remains a formidable challenge,
and despite centuries of research, only a handful of examples are
known to date.
In this article, we explore a distinct form of area-preserving
(symplectic) mappings derived from the stroboscopic Poincar\'e
cross-section of a kicked rotator --- an oscillator subjected to
an external force periodically switched on in short pulses.
The significance of this class of problems extends to various
applications in physics and mathematics, including particle
accelerators, crystallography, and studies of chaos.
Notably, Suris' theorem constrains the integrability within this
category of mappings, outlining potential scenarios with analytic
invariants of motion.
In this paper, we challenge the assumption of the analyticity of
the invariant by exploring piecewise linear transformations on a
torus ($\mathbb{T}^2$) and associated systems on the plane
($\mathbb{R}^2$), incorporating arithmetic quasiperiodicity and
discontinuities.
Introducing a new automated technique, we discovered previously
unknown scenarios featuring polygonal invariants that form perfect
tessellations and, moreover, fibrations of the plane/torus.
This work reveals a novel category of planar tilings characterized
by discrete symmetries that emerge from the invertibility of
transformations and are intrinsically linked to the presence of
integrability.
Our algorithm relies on the analysis of the Poincar\'e rotation
number and its piecewise monotonic nature for integrable cases,
contrasting with the noisy behavior in the case of chaos, thereby
allowing for clear separation.
Some of the newly discovered systems exhibit the peculiar behavior
of ``integrable diffusion,'' characterized by infinite and quasi-random
hopping between tiles while being confined to a set of invariant
segments.
Finally, through the implementation of a smoothening procedure,
all mappings can be generalized to quasi-integrable scenarios with
suppressed volume occupied by chaotic trajectories, thereby opening
doors to potential practical applications.
\noindent
Interactive simulations featuring visualizations of some of the
discovered maps are accessible at
\url{http://integrable-maps.streamlit.app}.
\end{abstract}



\maketitle

\section{Introduction}

Integrable systems, regarded as fundamental pillars in our
understanding of complex dynamics, manifest across diverse
disciplines such as celestial mechanics, hydrodynamics, nonlinear
optics, and condensed matter physics.
Within the domain of classical mechanics, archetypal integrable
Hamiltonians include well-recognized problems, such as the harmonic
oscillator, Kepler problem, Toda lattice, and the Euler top.
Considered mathematical ``gems'' in the realm of physics,
integrable systems are exceedingly rare, with only a handful of
textbook examples known to date
\cite{perelomov1990integrable,quispel1988integrable,quispel1989integrable}.

\newpage
In the study of Hamiltonian mechanics, symplectic mappings, such
as Poincar\'e or stroboscopic cross-sections derived from
higher-dimensional continuous systems, serve as essential bridges
between discrete and continuous dynamics.
On one hand, they serve as insightful tools for comprehending the
geometric nature of motion, and on the other, they can be
applied as numerical integrators for systems governed by
Hamilton's equations of motion.
Symplectic mappings are pervasive in physical applications and
naturally emerge in various contexts, including cyclic particle
accelerators~\cite{Meiss_RevModPhys.64.795,mcmillan1971problem,
PhysRevSTAB.11.114001}, Lagrange descriptions of fluid dynamics
and billiards, to name a few (e.g., see
\cite{veselov1991integrable}).

\newpage
Among all area-preserving transformations of the plane, our focus
is on a specific representation known as the McMillan-H\'enon (MH)
form of the map where dynamics is governed by a force function,
$f(q)$ (see~\cite{mcmillan1971problem,turaev2002polynomial,ZKhNArXiV}).
This form arises from the stroboscopic Poincar\'e cross-section
(time discretization) of a kicked rotator, a crucial toy model for
exploring classical and quantum chaos in non-autonomous Hamiltonian
systems.
It has a very clear analytical structure, and despite its seemingly
simple appearance, incorporates an exceptionally rich set of
scenarios, including (but not limited) to chaotic
H\'enon~\cite{henon1969numerical} and
Chirikov~\cite{chirikov1969research,chirikov1979universal},
and integrable McMillan mappings~\cite{mcmillan1971problem}.
Of particular significance is Suris' theorem
\cite{suris1989integrable}, which illuminates potential
integrable scenarios within this category.
Suris demonstrated that if an invariant of motion $\K[q,p]$ takes
the form of an analytic function, its possibilities are confined to
being regular, exponential, or trigonometric polynomials of degree
two.
This theorem serves as a robust constraint, shaping the landscape
of plausible integrable symplectic maps.

By challenging the assumption of analyticity of the invariant of
motion, one opens the possibility of discovering novel types of
integrable systems that go beyond the constraints
outlined in Suris' theorem.
Notable examples of mappings exhibiting linear dynamics, yet
possessing a nontrivial phase space structure, include the
Brown-Knuth map~\cite{brown1983,brown1985,brown1993}, McMillan
beheaded and two-headed ellipses~\cite{mcmillan1971problem},
and CNR maps~\cite{cairns2014piecewise,cairns2016conewise}.
In all instances, the force is a piecewise linear function, while
the integral of motion adopts a piecewise nature, characterized by
segments encompassing hyperbolas, ellipses, and/or line segments.
When all pieces are line segments, the phase space is fibrated
with polygons or polygonal chains, for stable and unstable cases
respectively.
In our earlier study~\cite{ZKhNArXiV}, we drew attention to the
existence of nonlinear systems with polygonal structure, which
led to the development of an automated search algorithm and the
subsequent discovery of a diverse set of new integrable maps.

Here, we present the next phase of our exploration, building upon
the foundation laid by previous results.
In this work, we examine piecewise linear forces comprised of an
infinite number of segments, adhering to a specific condition of
arithmetic quasiperiodicity.
In addition, we explore periodic piecewise linear forces that
allow for discontinuities, as well as a more general class of
systems constrained on a torus, which are related to planar
dynamics through a wrapping/unwrapping procedure.
These systems can be considered as a generalization of the famous
Bernoulli map \cite{schuster2006deterministic} (also known as the
dyadic transformation or bit-shift map) within planar symplectic
dynamics.
Moreover, not only did we uncover new large families of
previously unknown integrable systems but we also succeeded in
once again automating the search process.
This time, we employed a novel algorithm rooted in identifying the
rotation number ($\nu$) and utilizing analytical insights into the
piecewise monotonic dependence of $\nu$ on amplitude.
We would like to note that automated searches for integrable
systems are still in their early stages.
While some intriguing approaches, including machine learning, have
been recently developed (see ~\cite{krippendorf2021integrability}),
they are currently only capable of re-identifying systems that are
already known.

What adds intrigue to the discoveries is that, beyond contributing
to the arenas of integrable dynamics and algorithm theory, they
establish a connection with the problem of periodic tessellations
by polygons.
Such challenges, like the famous Euclidean plane tilings through
convex regular polygons, have captivated minds since antiquity.
The mathematical journey began with Kepler, who initiated a
systematic treatment in his monumental work ``Harmonices Mundi,''
and over time, this pursuit extended to encompass Archimedean,
aperiodic, Coxeter, and various other forms of tessellation.
Different types of tilings are characterized by specific rules
governing the permissible polygons and their arrangements, commonly
referred to as the ``tessellation rule.''

For all newly discovered systems, there exists a lattice constant
$L$ determining the translational symmetry of the invariant
$\K[q,p] = \K[q+m\,L,p+n\,L]$ for $m,\,n\in\mathbb{Z}$.
This symmetry is attributed to the periodic nature of $f(q)$.
Consequently, the constant level sets of $\K[q,p]$ corresponding
to isolating integrals of motion (separatrices) perfectly
tessellate the plane with polygons, while all other sets
contribute to a continuous fibration by polygons within all tiles
necessary for integrability.
In both instances, the rules governing tessellation and fibration
are intricately tied to the map's invertibility and its
decomposition into anti-area-preserving reflections
\cite{lewis1961reversible} (see Section~\ref{sec:Def}).
While there is a comprehensive understanding of certain types of
polygonal tessellations, in general, more sophisticated or aperiodic
patterns continue to be areas of active research
\cite{smith2023aperiodic,smith2023chiral,ZELLER2021102027,MANOLAS2022141},
and many key aspects have not achieved full clarity.
Perhaps, in this work, by establishing a connection between
polygonal tiling and integrability, we are adding a new, important
contribution regarding all possible systems of this type.

The paper is structured as follows.
In Section~\ref{sec:Def}, we introduce the MH form
of the map, provide definitions of force functions on both a torus
and a plane, and discuss various symmetries of the invariant.
Section~\ref{sec:Dynamics} offers clear examples illustrating how
integrable tessellations operate through the exploration of dynamic
regimes, including global mode-locking on the lattice and previously
unknown, to the best of our knowledge, regime of integrable diffusion.
In Section~\ref{sec:Chaos}, we introduce a new method for
identifying chaos with easy-to-follow examples, while the detailed
description of the automated search algorithm is left for
Appendix~\ref{secAPP:Algorithm}.
Section~\ref{sec:Results} reveals the main discoveries, and a
detailed analysis, including spectra and analytical expressions for
the rotation number, is presented at the end of the article in
Appendix~\ref{secAPP:Maps}.
Section~\ref{sec:Appl} discusses some possible applications
of the map, including a ``smoothening'' procedure that allows the
removal of all non-physical features, such as a lack of
differentiability or discontinuities, resulting in quasi-integrable
dynamics with suppressed chaotic behavior.

\newpage
\section{\label{sec:Def} Definitions and notations}

In this section, we begin by introducing key concepts such as
symplecticity, orbits, and integrability of the map, setting
the stage for a clear understanding of the system's dynamics.

$\bullet$
A mapping $\m:X \rightarrow X$ on a $2m$-dimensional manifold $X$
is {\it symplectic} if it satisfies the condition
\begin{equation}
\label{math:symplectic}
    \mathrm{J}_\m^\mathrm{T}\,\Omega\,\mathrm{J}_\m = \Omega,
\end{equation}
with
\[
\mathrm{J}_\m =
\begin{pmatrix}
\pd \mathbf{q}'/\pd \mathbf{q} & \pd \mathbf{q}'/\pd \mathbf{p} \\[0.2cm]
\pd \mathbf{p}'/\pd \mathbf{q} & \pd \mathbf{p}'/\pd \mathbf{p}
\end{pmatrix},
\qquad\qquad
\Omega =
\begin{pmatrix}
    0               & \mathrm{I}_m   \\[0.2cm]
    -\mathrm{I}_m   & 0
\end{pmatrix}.
\]
Here $\mathrm{J}_\m$ is the Jacobian matrix of $\m$, 
$\Omega$ is a $2m \times 2m$ nonsingular skew-symmetric matrix,
$\mathrm{I}_m$ is a $m \times m$ identity matrix,
a pair of $m$-dimensional variables, representing
configuration $\mathbf{q}$ and momenta $\mathbf{p}$, collectively
form a phase space coordinate $\zeta = (\mathbf{q},\mathbf{p})$,
and, $(')$ denotes the application of the map
\[
\zeta ' = (\mathbf{q}',\mathbf{p}') = \m (\mathbf{q},\mathbf{p}).
\]

In essence, symplecticity guarantees that the transformation
preserves the sum of areas projected on individual planes
$(q_i,p_i)$ within the phase space, emphasizing the geometric
structure of the dynamical system.
For mappings of the plane ($m=1$), the symplectic condition
(\ref{math:symplectic}) implies that $\det\mathrm{J}_\m = 1$,
indicating that the map preserves both area and orientation.

$\bullet$
The {\it orbit} of a map $\m$ is defined by a sequence of points,
$\zeta_k$, obtained through the repeated application of the map
to an initial condition, commonly referred to as the {\it seed},
$\zeta_0$:
\[
\left\{
    \zeta_0,\zeta_1,\zeta_2,\ldots
\right\},
\qquad\qquad
\zeta_k = \m(\zeta_{k-1}) = \m^k(\zeta_0).
\]
We say that the set of points
$\left\{\zeta_0,\zeta_1,\ldots,\zeta_{n-1}\right\}$
forms an $n$-{\it cycle} if
\[
    \m^n(\zeta_0) = \zeta_0.
\]
An $n$-cycle is a type of periodic orbit, and the smallest
positive integer $n$ for which the above condition holds is
called the {\it period of the cycle}.
If $n=1$, the map leaves the point unchanged after a single
iteration, and we refer to this state as a {\it fixed point},
indicated by  $\zeta_0 = \zeta_*$.
If $n=2$, it is a 2-cycle, and so on.

$\bullet$
A map of the plane is said to be {\it integrable} if there
exists a non-constant continuous function $\K[q,p]$, called
{\it integral} or {\it invariant of motion}, that remains conserved
upon the application of the map for all initial conditions, i.e.,
\begin{equation}
\label{math:integrability}
\forall\,(q,p)\in\mathbb{R}^2:
\qquad
\K[q',p'] - \K[q,p] = 0.
\end{equation}
This implies that the orbits originating from $\zeta_0$ lie on a
constant level of $\K = \K(\zeta_0)$, whether it manifests as a
singular point, a set of points, or as a single (or collection)
of closed/open curve(s).

\newpage
\subsection{McMillan-H\'enon form of the map}

This article focuses on a particular representation of the
symplectic mapping of the plane referred as the McMillan-H\'enon
(MH) form.
A detailed overview can be found in
\cite{mcmillan1971problem,turaev2002polynomial,ZKhNArXiV},
while here we briefly list some of its relevant properties.
The transformation and its inverse are explicitly expressed as:
\begin{equation}
\label{math:MTform}    
\begin{array}{ll}
\m:         & q' = p,        \\[0.25cm]
                        & p' =-q + f(p),
\end{array}
\quad\,\,\,\,
\begin{array}{ll}
\m^{-1}:   & q' =-p + f(q), \\[0.25cm]
                                & p' = q,
\end{array}
\end{equation}
where $f(q)$ is called the {\it force function}.
This map maintains symplecticity for any $f(q)$, offering a
framework for various well-known examples.
These include the extensively studied Chirikov standard map
\cite{chirikov1969research,chirikov1979universal},
the area-preserving H\'enon quadratic map~\cite{henon1969numerical},
or a collection of integrable mappings discovered by
E. McMillan~\cite{mcmillan1971problem} and
Y. Suris~\cite{suris1989integrable}.


The map in MH form can be written as a composition
$\m = \mathcal{F} \circ \mathrm{Rot}(\pi/2)$,
where $\mathrm{Rot}(\theta)$ is a rotation about the origin
\[
    \mathrm{Rot}(\theta):\quad
    \begin{bmatrix} q'\\ p' \end{bmatrix} =
    \begin{bmatrix}
        \cos\theta & \sin\theta\\
       -\sin\theta & \cos\theta
    \end{bmatrix}
    \begin{bmatrix} q \\ p  \end{bmatrix}
\]
followed by a thin lens transformation
\[
    \mathcal{F}:\quad
    \begin{bmatrix} q'\\ p'  \end{bmatrix} =
    \begin{bmatrix} q \\ p   \end{bmatrix} +
    \begin{bmatrix} 0 \\ f(q)\end{bmatrix}.
\]
This breakdown unveils the physical nature of the system, depicting
a linear oscillator subject to periodic kicks in time, influenced by
the force $f(q)$.
In specific applications, a model accelerator featuring one degree
of freedom and comprising a linear optics insert in tandem with a
thin nonlinear lens (e.g., sextupole, octupole or RF-station), can
be redefined in the MH form.

Alternatively, $\m$ can be decomposed as
$\mathcal{G}\circ\mathrm{Ref}(\pi/4)$,
where $\mathrm{Ref}(\theta)$ is a reflection about a line passing
through the origin at an angle $\theta$ with the $q$-axis
\[
\mathrm{Ref}(\theta):\quad
    \begin{bmatrix} q' \\ p' \end{bmatrix} =
    \begin{bmatrix}
        \cos2\theta & \sin2\theta\\
        \sin2\theta &-\cos2\theta
    \end{bmatrix}
    \begin{bmatrix} q  \\ p  \end{bmatrix},
\]
and $\mathcal{G}$ is a nonlinear vertical reflection
\[
\mathcal{G}:\quad 
    \begin{bmatrix} q'\\ p'  \end{bmatrix} =
    \begin{bmatrix} q \\-p   \end{bmatrix} +
    \begin{bmatrix} 0 \\ f(q)\end{bmatrix}.
\]
Both, $\mathrm{Ref}(\theta)$ and $\mathcal{G}$, are
{\it anti-area-preserving involutions}, which means that a double
application of the map is equivalent to identity transformation
$\mathrm{Ref}^2(\theta)=\mathcal{G}^2=\mathrm{I}_2$ and the
determinant of their Jacobians is equal to $-1$.
Each reflection possesses a line of fixed points:
\[
l_1:\,\, p = q      \,\,\,\text{for}\,\,\,\mathrm{Ref}(\pi/4),
\qquad
l_2:\,\, p = f(q)/2 \,\,\,\text{for}\,\,\,\mathcal{G},
\]
designated as the {\it first} and {\it second symmetry lines}
\cite{lewis1961reversible,mcmillan1971problem}.

\subsection{Force functions with patterns of regularity}

In~\cite{ZKhNArXiV}, we introduced numerous nonlinear integrable
mappings with polygon invariants and corresponding piecewise
linear force functions, denoted here as $f_\mathrm{p.l.}(q)$.
To further expand upon this result, one can pursue a straightforward
approach by exploring forces that incorporate an increasing number
of segments.
Alternatively, we can take a different path to extend our findings
and examine dynamics with piecewise linear forces on a torus
$\zeta \in \mathbb{T}^2$, i.e., when the equations of
motion~(\ref{math:MTform}) are considered modulo $L$, where $L$ is
referred as the {\it lattice constant}.

\begin{figure}[t!]\centering
\includegraphics[width=\columnwidth]{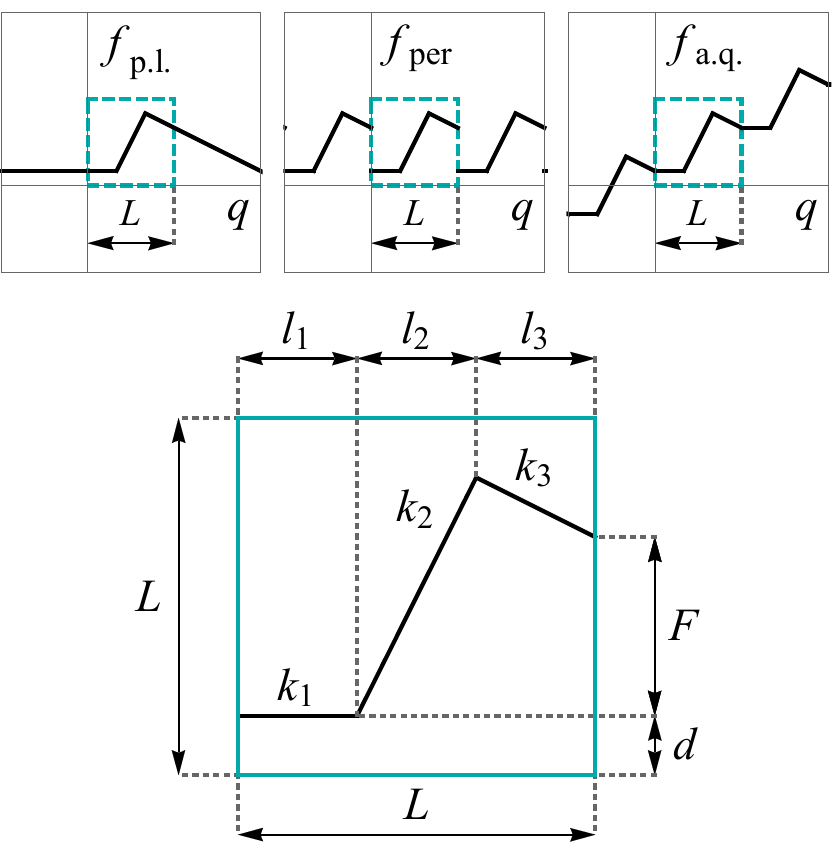}
\vspace{-0.5cm}
\caption{\label{fig:ForceFunctions}
Top row illustrates a regular piecewise linear force function
composed of three segments (left), along with its periodic (middle)
and arithmetically quasiperiodic (right) unwrapping from the torus
(mod $L$), depicted with a cyan square.
The bottom plot displays the corresponding force on a torus,
providing visual references for the introduced notations.
}\vspace{-0.5cm}
\end{figure}

Furthermore, we return our study to the planar dynamics
$\zeta\in\mathbb{R}^2$, through two possible ``unwrappings''.
The first one,
\begin{equation}
\label{math:ftor}    
    f_\mathrm{per}(q) = f_\mathrm{p.l.}(q\!\!\!\mod L) \mod L
\end{equation}
results in force that consists of an infinite number of segments,
strictly periodic $f_\mathrm{per}(q) = f_\mathrm{per}(q + L)$,
and might include discontinuities, that can be seen as the inclusion
of additional segments with their lengths approaching zero and slopes
tending towards infinity.
Otherwise, we can impose the {\it arithmetical quasiperiodicity},
which implies that
\begin{equation}
\label{math:ArQuas}    
    \forall\,q:\quad
        f_\mathrm{a.q.}(q+L) = f_\mathrm{a.q.}(q) + F
\end{equation}
where
\[
    F = f_\mathrm{p.l.}(L) - f_\mathrm{p.l.}(0) = \const.
\]
As evident, a function with arithmetic quasiperiodicity can be
expressed as the sum of a purely periodic and linear functions
\begin{equation}
\label{math:ArQuas2}
\quad\,\,
    f_\mathrm{a.q.}(q) = d+k_0\,q + f_\mathrm{per}(q),
\end{equation}
where $k_0=F/L$ provides the average slope of $f_\mathrm{a.q.}(q)$.
In this article, we will utilize notations that define the force
function as $\left[\{k_1,l_1\},\ldots,\{k_n,l_n\};d\right]$,
where each of the $n$ tuples $\{k_j,l_j\}$ represents the slope
and length of the $j$-th segment for $q\in[0,L]$, followed by the
value of the vertical {\it shift parameter}, $d = f_\mathrm{p.l.}(0)$.
The top row of Fig.~\ref{fig:ForceFunctions} provides illustrative
examples of piecewise linear forces on a plane: regular, periodic
with discontinuities, and a function with arithmetic quasiperiodicity.
Their representation on a torus (mod $L$) is shown at the bottom,
offering visual cues for the introduced notations.

{\it
Note, it is crucial to avoid confusing the concept of
arithmetic quasiperiodicity with the idea of {\bf quasiperiodic
motion} in the context of a symplectic map of the plane.
In the latter case, the trajectory of the system forms a dense
curve (or collection of curves) in phase space, and although it
never exactly repeats, it approaches previously visited points
arbitrarily closely.
In other words, quasiperiodicity of oscillations implies that
the rotation number is irrational, unlike in periodic motion,
where the trajectory repeats after a fixed number of iterations
with a rational rotation number.}

The form of the map and the repetitive nature of the force
function give rise to several sources of redundancy in the
parameter space.
These redundancies are as follows:

$\bullet$ Each map has a corresponding ``twin'' map, which can
be obtained by using the transformation
$f(q)\rightarrow -f(-q)$.
Geometrically, this is equivalent to rotating the phase space
by an angle of $\pi$, i.e. $(q,p) \rightarrow -(q,p)$.

$\bullet$ Cyclic permutation of segments in the force function
results in the same map, provided that the shift parameter
$d$ is adjusted to maintain the relative location of the
fixed point, $q_*$.

$\bullet$ For any given set of parameters, there exists a set
of equidistant values of the parameter $d$, expressed as
$d = d_0 + m\,(2\,L-F)$, where $m \in \mathbb{Z}$ is an
integer.
This generates an equivalent dynamical system with the fixed
point $q_* \rightarrow q_* + m\,L$, owing to the translational
symmetry of the invariant.
    
$\bullet$ Lastly, due to the linear nature of each segment,
the rescaling of dynamical variables
$(q,p) \rightarrow \epsilon\,(q,p)$ leaves the form of the map
invariant.
This rescaling allows us to choose one of the segments
(or, e.g., $L$ or $d$) to have a length equal to one.

\noindent
In order to avoid reporting duplicated results, we will eliminate
any redundant solutions and scaling parameters whenever possible
during our analysis.
This will ensure that we focus on distinct and meaningful outcomes
in our exploration of the parameter space.

\subsection{Symmetries of the invariant}

Primarily, for the mapping in the MH form (\ref{math:MTform}), any invariant level
set (even amid chaos and the absence of a global invariant for
all initial conditions) remains unchanged not only under $\m$,
but also when subjected to either one of the reflections,
$\mathrm{Ref}(\pi/4)$ and $\mathcal{G}$, respectively
\begin{equation}
\label{math:integrability2}
    \K[q,p] = \K[p,q]
    \quad\text{ and }\quad
    \K[q,p] = \K[q,-p+f(q)].
\end{equation}
If the set of Eqs.~(\ref{math:integrability2}) is satisfied for
all points in the plane/torus, the map $\m$ is integrable.
This offers a straightforward analytical test to verify
integrability by confirming that every invariant set is symmetric
with respect to the main diagonal $p = q$ and consists of parts that
are vertically equidistant from $p=f(q)/2$.

At the same time, when unwrapping the integrable system from the
torus to the Euclidean plane, we introduce the {\it lattice
translation symmetry} of the invariant
\[
    \K[q,p] = \K[q+m\,L,p+n\,L],
    \qquad
    m,n\in\,\mathbb{Z},
\]
which arises from the patterns of regularity in both
$f_\mathrm{per}(q)$ and $f_\mathrm{a.q.}(q)$.
Consequently, the constant level sets of $\K[q,p]$, corresponding
to isolating integrals of motion/separatrices on a torus, now
perfectly tessellate the plane with polygons.
Using terminology from famous Archimedean tilings, we designate the
tiling as {\it regular} if the phase space is tessellated with a
single type of polygon separatrices, and as {\it semi-regular} when
the net of separatrices consists of multiple polygon types while
maintaining a coherent pattern across the entire plane.
Simultaneously, all other invariant level sets contribute to a
continuous fibration by polygons within all tiles necessary for
integrability.
In both instances, the governing tessellation/fibration rules are
provided by Eq.~(\ref{math:integrability2}) and are satisfied
$\forall\,\zeta\in\mathbb{R}^2$ ($\in\mathbb{T}^2$).

Finally, some of the newly discovered systems possess an
additional discrete translational symmetry of the invariant:
it is possible that, for a given tessellation by polygons,
Eqs.~(\ref{math:integrability2}) can be satisfied for some
values of $d$ that differ from $d_0 + m\,(2\,L-F)$, where $m$
is an integer.
This symmetry will be called the {\it configuration symmetry}
because it is associated with the relative position of the fixed
point, which for the mapping in MH form is provided by the
intersection of symmetry lines, $l_1$ and $l_2$:
\[
    q_*=p_*:\qquad q_* = f(q_*)/2.
\]
For instance, in a regular tiling, the fixed point can be located
either inside one of the tiles, which we refer to as the
{\it central cell} (or cc for brevity), or it can be placed into
the node/vertex of the separatrix net.
In the latter case, we say that the map has a {\it central node}
configuration (or cn for short).

It is worth noting that owing to the lattice translation symmetry
of the invariant, for every integrable map with $f_\mathrm{per}(q)$
or $f_\mathrm{a.q.}(q)$, we can define a corresponding integrable
system on a torus.
However, the converse guarantees integrability only for purely
periodic functions.
In other words, the ``unwrapping'' from the torus while attempting
to enforce arithmetical quasiperiodicity can result in the second
symmetry line $f_\mathrm{a.q.}(q)/2$ that is incompatible with the
lattice.

\begin{figure}[t!]\centering
\includegraphics[width=\columnwidth]{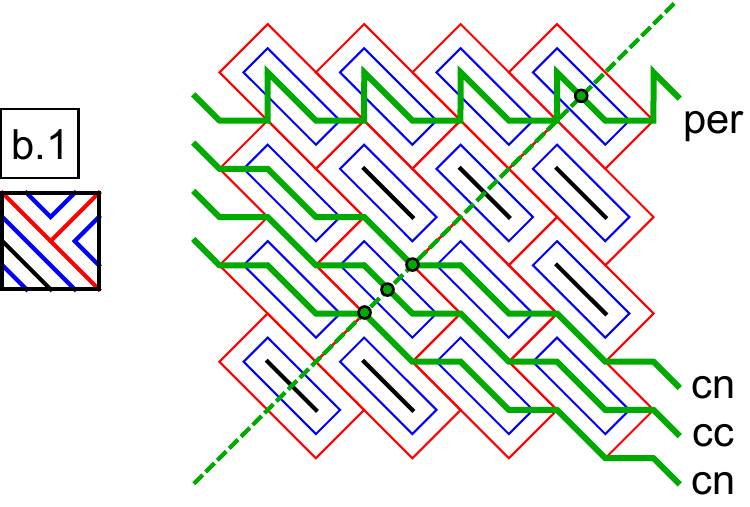}
\caption{\label{fig:Unwrap}
    Phase space portrait for the map $\m_\mathrm{b1}^\mathrm{tor}$
    along with the corresponding unwrapping of the invariant to
    $\mathbb{R}^2$.
    Various second symmetry lines illustrate possible configurations
    of the map: periodic force with discontinuities (per), or,
    arithmetically quasiperiodic force allowing for the central cell
    (cc) and central nodes (cn).\\
    {\bf
    Throughout this paper, we employ the following color scheme to
    represent invariant level sets in phase space: red corresponds
    to isolating invariants/separatrices, while blue and black
    indicate phase space orbits within layers with nonlinear
    (amplitude-dependent) and linear (mode-locked) dynamics,
    respectively.
    The solid and dashed green lines in all diagrams correspond
    to the second, $p=f(q)/2$, and first symmetry lines, $p=q$.
    }
}       
\end{figure}

To effectively illustrate the concepts discussed, we use a simple
integrable map on a torus with a 2-piece force function defined as:
\[
\m_\mathrm{b1}^\mathrm{tor}:\quad
    f_\mathrm{tor}(q) = [\{-2,L/2\},\{0,L/2\};2\,L].
\]
Figure~\ref{fig:Unwrap} displays the phase space portrait on the
torus (left) and the corresponding ``unwrapping'' of the invariant
to the Euclidean plane (right).
Organizing the phase space portrait (mod $L$) on a regular square
grid in $\mathbb{R}^2$ reveals a regular tessellation.
Each tile contains an invariant line segment with 2-cycles
(depicted in black), while the remaining phase space is filled with
concentric polygons (depicted in blue), bounded by the rectangular
separatrix (depicted in red).
This tessellation not only aligns with periodic unwrapping, as
intended in its construction, but also accommodates unwrapping with
arithmetic quasiperiodicity.
Additionally, the system allows for different fixed point locations,
either inside one of the tiles (cc) or at the separatrix nodes (cn),
owing to the presence of configuration symmetry.

\section{\label{sec:Dynamics} Exploring Dynamical Regimes}

\begin{figure}[b!]\centering
\includegraphics[width=0.98\columnwidth]{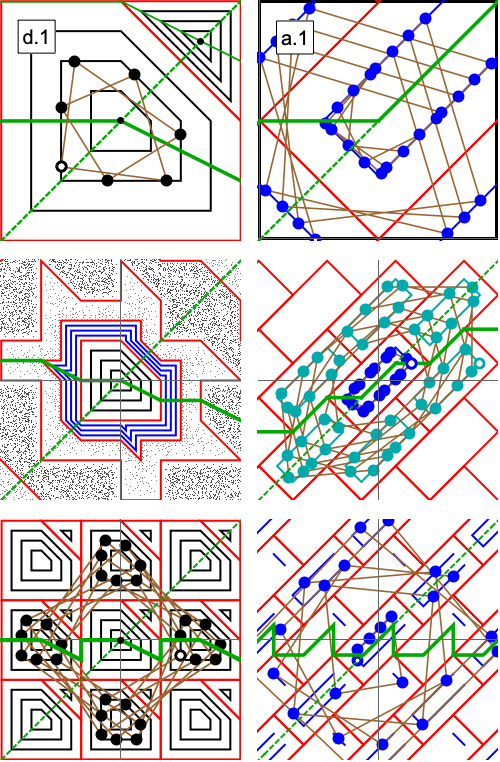}
\vspace{-0.3cm}
\caption{\label{fig:Dynamics}
    Phase space portraits for the mappings on a torus,
    $\m_{\mathrm{d}1}^\mathrm{tor}$ and
    $\m_{\mathrm{a}1}^\mathrm{tor}$ (top row),
    along with their corresponding unwrapping with arithmetical
    quasiperiodicity (middle row) and periodic unwrapping with
    discontinuities (bottom row).
    The left plot in the middle, for the case (d.1), is non-integrable;
    chaotic trajectories are depicted with points in varying shades
    of gray.
    All other systems maintain integrability, showcasing sample
    orbits represented by series of connected dots.
}       
\end{figure}

In order to describe dynamics, we consider two additional systems
that are integrable on a torus, each associated with corresponding
forces derived from piecewise linear functions consisting of two
segments:
\[
\begin{array}{ll}
\ds\m_\mathrm{d1}^\mathrm{tor}:\quad &\ds
    f_\mathrm{tor}(q) = [\{0,L/2\},\{-1,L/2\};L],   \\[0.35cm]
\ds\m_\mathrm{a1}^\mathrm{tor}:\quad &\ds
    f_\mathrm{tor}(q) = [\{0,L/2\},\{2,L/2\};L].
\end{array}
\]
Figure~\ref{fig:Dynamics} presents phase space portraits on a torus
(top row) along with their possible unwrappings to the plane.
For each integrable scenario, we provide sample orbits represented
by series of connected points.
In the case of mappings involving chaotic behavior, scattered dots
obtained through numerical iterations are used instead.

\vspace{-1.95cm}

\newpage
\subsection{Dynamics (mod $L$)}

\vspace{-0.2cm}
We begin our exploration by addressing dynamics modulo $L$.
While, in general, the main aspects closely resemble maps with
regular piecewise linear forces $f_\mathrm{p.l.}$ \cite{ZKhNArXiV},
subtle nuances exist.
Primarily, for regular mappings on the plane, the orbits can be
either stable (polygons) or unstable (simple polygonal chains).
However, due to the finite and compact nature of the torus surface
in contrast to the infinite and unbounded plane, we refrain from
explicitly considering stability.
Instead, we discern between {\it oscillations}, where the
trajectory rounds the fixed point, tracing out closed-loop paths
(as can be seen for the map (d.1)), and, {\it drift} or {\it flow},
where the orbit ``weaves'' and ``loops'' around the torus upon
reaching its border.
One can draw an analogy to the two regimes of motion observed in a
classical pendulum.
Nonlinear flow implies trajectories that are more intricate than
simple circular or closed-loop paths, as they can traverse
different regions of the torus, as seen in example (a.1) in
Fig.~\ref{fig:Dynamics}.

Moreover, concerning planar dynamics, we categorize stable orbits
as either purely periodic ($n$-cycles) or quasiperiodic, contingent
on whether the rotation number is rational or irrational.
This concept extends to both types of motion on a torus, where
the term {\it winding number} is sometimes employed to distinguish
drift from oscillations with rotation number.
Mapping (d.1), being degenerate, yields all orbits as 7- or
3-cycles, while the dynamics of map (a.1) is nonlinear,
leading to a dense coverage of most orbits across their invariant
level sets.

\vspace{-0.4cm}
\subsection{\label{sbs:GlobMode} Global mode-locking}

\vspace{-0.2cm}
Next, we turn our attention to planar dynamics with continuous
forces exhibiting arithmetical quasiperiodicity.
As previously noted, unwrapping the system in this way while
maintaining integrability may be impossible, as illustrated by
the left plot in the middle of Fig.~\ref{fig:Dynamics}, where
the presence of chaotic orbits is evident.
On the contrary, mapping (a.1) facilitates integrability for
$f_\mathrm{a.q.}(q)$ and generates a regular tessellation with
a central cell, similar to the previously examined map (b.1).

While all tiles of the same type are geometrically identical, the
dynamics on similar level sets of the invariant vary from one
cell of periodicity to another.
Foremost, there exists a unique tile (or a unique node in the case
of the cn configuration) containing a fixed point.
Within the cc, the dynamics revolve around the fixed point, akin
to the scenario observed in regular polygon maps.

Furthermore, all initial conditions on the separatrix isolating
cc correspond to $P$-fold periodic orbits ($P$-cycles), causing
the non-central tiles to form chains of $P$ islands.
As a result, initial conditions from non-central tiles exhibit
{\it global mode locking} on the lattice with a rational rotation
number defined by:
\begin{equation}
\label{math:ModeLNu}
    \nu_P = \arccos[k_0/2]/(2\,\pi) =
        R/P,\qquad R\in\mathbb{Z}.
\end{equation}
Here, we refer to $P \in \mathbb{N}$ and $\nu_P \in \mathbb{Q}$ as
the {\it global period} and {\it global rotation number} of the map.

\newpage
\begin{figure}[t!]\centering
\includegraphics[width=\columnwidth]{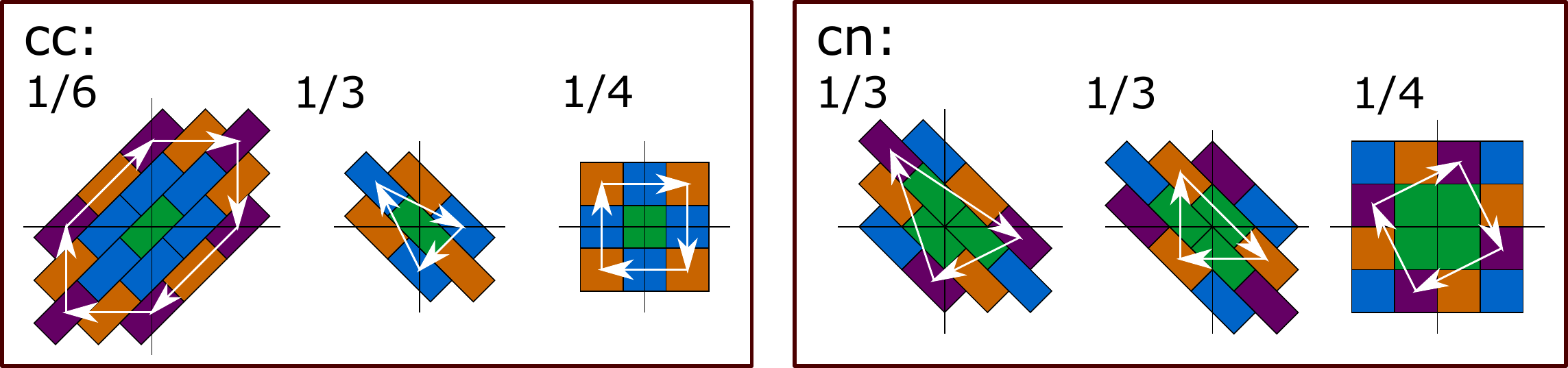}
\caption{\label{fig:CCCN}
    Schematic illustration of the global mode-locking around the
    central cell [cc] or central node [cn] for different values of
    $\nu_P$.
    Tiles of the same color are grouped in chains of islands, and
    the arrows indicate the order in which particular tiles are
    visited.
}\vspace{-0.5cm}
\end{figure}

Figure~\ref{fig:CCCN} schematically depicts the arrangement in
which tiles are grouped into islands, with $P$ values of 6, 3,
and 4.
The occurrence of these specific periods is not arbitrary and is
determined by the integer values of
\[
    k_0 = 1,\,\,-1\,\,\text{and}\,\,0.
\]
These values are in accordance with Crystallographic theorem
\cite{cairns2014piecewise,cairns2016conewise,ZKhNArXiV} and
naturally emerge when dealing with integer coefficients.
It remains an open question whether it is possible to organize
segments of the force function to produce a tessellation with $P$
values different from the aforementioned ones, requiring
$k_0\in\mathbb{Q}$.
A more detailed illustration is presented in the right plot in the
middle of Fig.~\ref{fig:Dynamics}, providing an example of two
orbits generated from initial conditions separated by one lattice
period.
The blue and cyan dots/polygons correspond to initial conditions
within the central cell $(q_0, p_0)\in\mathrm{cc}$ and $(q_0+L, p_0)$, respectively.
In the latter case, the trajectory sequentially visits six islands
under iterations and traces out a collection of invariant tori,
see additional plot in Fig.~\ref{fig:IntDif}.

\vspace{-0.2cm}
\subsection{\label{sbs:IntDiff} Integrable diffusion}

\vspace{-0.2cm}
Here, we proceed to the last scenario of periodic functions
resulting from simple unwrapping from the torus,
Eq.~(\ref{math:ftor}).
While the continuity of $f_\mathrm{a.q.}(q)$ from the previous
subsection guarantees the continuity of the invariant of motion,
allowing global mode-locking, the inclusion of discontinuities
can lead to significant differences.

The left plot at the bottom row of Fig.~\ref{fig:Dynamics}
provides a corresponding unwrapping for the map (d.1), showing
another sample orbit, this time departing from a non-central cell
located one lattice period to the right.
As one would expect, the sample orbit is mode-locked to $1/4$,
since the average slope $k_0$ is equal to zero, and it hops
between four islands, forming $7 \times 4$-cycles (or
$3 \times 4$-cycles if departing from a non-central triangle).

\begin{figure}[b!]\centering
\includegraphics[width=\columnwidth]{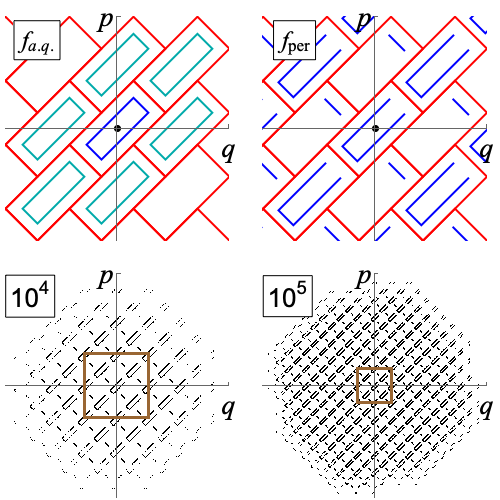}
\vspace{-0.5cm}
\caption{\label{fig:IntDif}
    Top row displays the replication of invariant level sets traced
    out by sample trajectories for the unwrappings of (a.1),
    Fig.~\ref{fig:Dynamics}.
    The bottom row illustrates integrable diffusion for the top-right
    plot, depicting $10^4$ and $10^5$ iterations obtained by tracking,
    with a brown square having a side length of $2\,L$ for reference.
    }
\end{figure}

While unwrapping from the torus might appear somewhat trivial,
involving the creation of new invariant level sets through simple translations, unexpected behavior arises in the case of mapping
(a.1), illustrated at the bottom right corner of
Fig.~\ref{fig:Dynamics}.
Surprisingly, after several iterations, the orbit originating
inside the central cell escapes and begins to hop between tiles
in a rather intricate manner.
However, instead of exhibiting a ``random'' trajectory, the orbit
remains on a specific level set, as further demonstrated in
Fig.~\ref{fig:IntDif}.

Comparing this case to the previously considered unwrapping with
arithmetic quasiperiodicity, as shown in Fig.~\ref{fig:IntDif},
reveals that instead of a finite number of closed rectangles,
we now have infinitely many sets of rectangles split into two
pieces.
The two plots at the bottom provide $10^4$ and $10^5$ iterations
of the same sample orbit as in Fig.~\ref{fig:Dynamics} but on a
different scale, clearly depicting diffusion.
Notice how the progression in time unveils more and more segments
of the invariant, aligning with the observed pattern.
We will refer to this behavior as ``integrable diffusion.''

Now, we can raise the question: ``Why do cases (a.1) and (d.1)
exhibit such different behavior when considered on the plane?''
The resolution to this puzzle lies in how the network of
separatrices (shown in red) is organized within the square
representing the torus:
in case (d.1), separatrices form closed loops and are confined
within $\mathbb{T}^2$, thus providing isolating invariants when
the system is unwrapped.
On the other hand, for case (a.1), the boundary of the torus cuts
the separatrix, creating gaps in isolating invariants and allowing
for diffusion.
In fact, one can observe that both behaviors fit into the larger
picture where we had stable and unstable orbits for regular
piecewise mappings of the plane, then the corresponding pair of
rotations around a fixed point (oscillations) and flow (drift or
winding) on a torus, and finally, global mode locking and
integrable diffusion when force exhibits patterns of regularity.

\vspace{-0.2cm}

\newpage
\section{\label{sec:Chaos} Chaos detection}

After demonstrating the possibility of constructing nonlinear
integrable mappings allowing tessellation, a new question
emerged:
``How can we find more systems and automate the search process''?
In~\cite{ZKhNArXiV}, where $f_\mathrm{p.l.}(q)$ represented a
standard piecewise linear force function, our approach relied on
the assumption that the phase space should be fibrated with
concentric polygons.
By organizing points of an individual orbit based on polar
angles and counting the number of vertices, we could identify
non-integrable systems where the polygons produced after
rearranging chaotic trajectories had an increasing number of
vertices that did not converge with the number of iterations.

Unfortunately, as mentioned in~\cite{ZKhNArXiV}, this algorithm
fails when the phase space involves islands with nonlinear
dynamics, i.e., in the presence of global mode-locking:
arranging the points of the corresponding orbit by polar angle
becomes problematic, as $\K(r,\theta)=\const$ transforms into a
multiple-valued function of $\theta$ defined on disjoint intervals,
as can be seen in top left plot in Fig.~\ref{fig:IntDif}.

To address this issue, we decided to rely on intrinsic dynamical
variables rather than analyzing the structural aspects of the phase
space.
We first recall the canonical form of an integrable system
\cite{arnold1968ergodic,arnol2013mathematical,lichtenberg2013regular}:
\[
\begin{array}{l}
    J' = J,                     \\[0.25cm]
    \psi' = \psi + 2\,\pi\,\nu(J),
\end{array}
\qquad\qquad
\begin{array}{l}
    J_k = J_0,                  \\[0.25cm]
    \psi_k = \psi_0 + 2\,\pi\,k\,\nu(J_0),
\end{array}
\]
where $J_k$ and $\psi_k$ are the action and angle variables at
the $k$-th iteration, and $\nu$ denotes the {\it rotation number}
which captures the average angular increment of the angle variable
during a single iteration of the map.
$\nu$ remains invariant regardless of the mapping’s representation,
and for instance, in $(q,p)$-coordinates, it can be analytically
expressed using Danilov’s theorem
\cite{zolkin2017rotation,nagaitsev2020betatron,mitchell2021extracting}
or as a limit:
\begin{equation}
\label{math:rotationNumber}
\nu =   \frac{1}{2\,\pi}\,\lim_{N\rightarrow\infty}
        \frac{\m^N(\theta_0)-\theta_0}{N}
    =   \frac{1}{2\,\pi}\,\lim_{N\rightarrow\infty}
        \frac{\theta_N-\theta_0}{N},
\end{equation}
where $\theta_k = \arctan[(p_k-p_*)/(q_k-q_*)]$ is the polar
angle measured with respect to the fixed point.
For chaotic trajectories, this limit is non-existent due to
the erratic and unpredictable behavior of the trajectory's
angle, which prevents the establishment of a well-defined
average angular increment.

Furthermore, it is important to recognize that the phase space is
effectively segmented into layers defined by the pieces
of the force function and ``demarcated'' by separatrices.
The utilization of Danilov's theorem uncovers that within each
layer the Poincar\'e rotation number must be of the
form~\cite{ZKhNArXiV}
\begin{equation}
\label{math:rotationJ}
    \nu(J) = \frac{a_0+a_1\,J}{b_0+b_1\,J}
\end{equation}
where $a_{0,1}$ and $b_{0,1}$ represent specific constants.

\newpage
Observing that Eq.~(\ref{math:rotationJ}) constitutes a monotonic
function of amplitude enables us to distinguish integrable layers
from those involving chaotic trajectories.
In the latter scenario, the rotation number will not exhibit
monotonic behavior, but rather showcase discontinuous fluctuations
during numerical experiments.
To provide a visual representation of the concept underlying this
new approach, we employ three mappings characterized by force
functions $f_\mathrm{a.q.}(q)$ derived from the subsequent
parameter sets:
\[
\begin{array}{ll}
    \left[ \{ 0;1\},\{-1;1\},\{-2;1\},\{-1;1\};-5  \right] &\!\!
    (\text{integrable map})      \\[0.25cm]
    \left[ \{-2;1\},\{ 1;2\}                  ; 1  \right] &\!\!
    (\text{chaos in cell})       \\[0.25cm]
    \left[ \{ 1;1\},\{ 0;2\},\{-1;1\},\{ 0;1\};1/2 \right] &\!\!
    (\text{chaotic diffusion})
\end{array}
\]
To analyze these dynamical systems, we will utilize the rotation
number as a function of the horizontal distance from the initial
condition to the fixed point, denoted by $s = q_0 - q_*$, along
both symmetry lines:
\[
    \nu_1 = \left. \nu(s) \right|_{\zeta_0 = (q_0,q_0)}
    \quad\mathrm{and}\quad
    \nu_2 = \left. \nu(s) \right|_{\zeta_0 = (q_0,f(q_0)/2)}.
\]
Both spectra are evaluated within two lattice periods, as indicated
on the phase portraits by a dashed cyan square.

\vspace{-0.2cm}
\subsection{Integrable map}

\begin{figure}[b!]\centering
\includegraphics[width=\columnwidth]{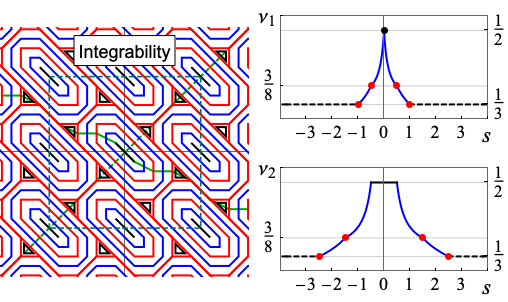}
\vspace{-0.8cm}
\caption{\label{fig:Example1}
    Phase space portrait and spectra along symmetry lines for the
    case with integrable dynamics.
    }
\end{figure}

First, let's explore a more intricate example of a nonlinear
integrable system featuring a semi-regular tiling pattern,
illustrated in Fig.~\ref{fig:Example1}.
Around the fixed point, there are three layers: a line segment
with 2-cycles (black line), succeeded by two layers consisting
of concentric polygons with nonlinear dynamics (depicted by blue
contours separated by red separatrices).
The cc is replicated throughout the phase plane, and the gaps
between the bounding octagons are filled with invariant triangles.
The average slope $k_0$ of the force $f_\mathrm{a.q}(q)$ is
equal to $-1$, leading to the organization of non-central tiles
into chains of three islands, with a corresponding global rotation
number of $\nu_P = 1/3$.
As anticipated, the spectra display piecewise monotonic behavior
inside the cc, indicated by corresponding colors, with global mode-locking
outside matching $\nu_P$, represented by dashed black lines.
\vspace{-1cm}

\newpage
\begin{figure}[t!]\centering
\includegraphics[width=\columnwidth]{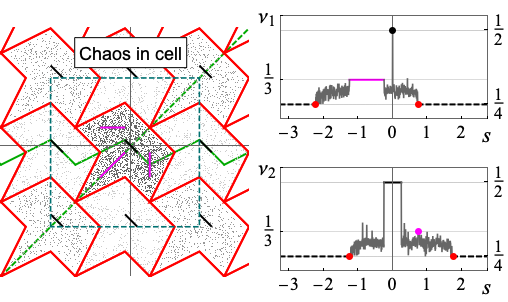}
\vspace{-0.8cm}
\caption{\label{fig:Example2}
    Phase space portrait and spectra along symmetry lines for the
    case of ``chaos in cell.''
    }\vspace{-0.2cm}
\end{figure}

\subsection{Chaos in cell}

\vspace{-0.2cm}
Moving on to the next example shown in Fig.~\ref{fig:Example2},
we delve into a scenario that addresses the notion of ``moderate
perturbation'', primarily affecting the small and intermediate
amplitudes of the system.
To illustrate this situation, we have chosen an extreme case where
the network of separatrices is the only set of invariant tori that
remains intact.
These isolating invariants wield a significant influence on the
system's dynamics by effectively segregating cells of periodicity.
The average slope is $k_0 = 0$, resulting in a global mode locking
of $1/4$.
Consequently, all points located on the separatrix possess a period
equal to $P=4$.

Within the central cell of periodicity, there are two additional
invariant structures of unit measure that emerge: a line segment
featuring 2-cycles around the fixed point (depicted as a black
line) and a chain of three islands composed of line segments
(highlighted in magenta).
The clusters of dots, varying in shades of gray, represent three
chaotic trajectories that are either ``locked'' within the central
cell or enclosed by the surrounding cells grouped in fours
(as illustrated in Fig.~\ref{fig:CCCN}).
This particular scenario is termed as the {\it chaos in cell},
denoting the presence of chaotic behavior within a confined region.

When examining the corresponding spectra along the symmetry lines,
it becomes evident that, for nearly all initial conditions within
the central cell (excluding non-isolated 2- and 3-cycles), the
rotation numbers exhibit chaotic signatures akin to noise.
Outside the central cell, as anticipated, the rotation number
averages out and becomes mode-locked to $1/4$.
Red dots represent the locations of corresponding separatrices.

\vspace{-0.25cm}
\subsection{Chaotic diffusion}

\begin{figure}[h!]\centering
\includegraphics[width=\linewidth]{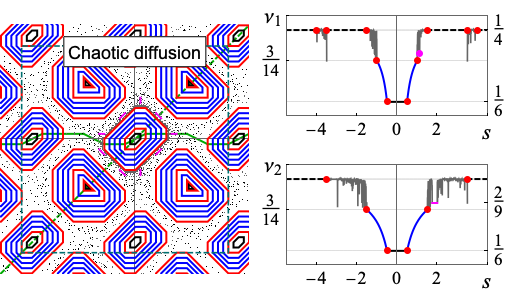}
\vspace{-0.75cm}
\caption{\label{fig:Example3}
    Phase space portrait and spectra along symmetry lines for the
    case of ``chaotic diffusion.''
    }
\end{figure}

\vspace{-0.15cm}
Finally, we consider a scenario where perturbation primarily
impacts the larger amplitudes of the system (within the lattice
period), causing a complete disruption of the invariant isolating
structures, as illustrated in Fig.~\ref{fig:Example3}.
In this selected example, the phase space consists of two integrable
fragments following a semi-regular pattern and separated by a 
``chaotic sea.''
The central cell (and corresponding tiles) exhibits two distinct
layers surrounding the fixed point: concentric hexagons (depicted
in black) and octagons (highlighted in blue).
The second integrable fragment is composed of three layers made of
triangles, pentagons, and octagons.
All layers shown with contours in black experience degeneracy,
with every initial condition being periodic, while contours in
blue represent layers with amplitude-dependent dynamics,
$\nu(s)\neq\const$.

\begin{figure}[t]\centering
\includegraphics[width=\linewidth]{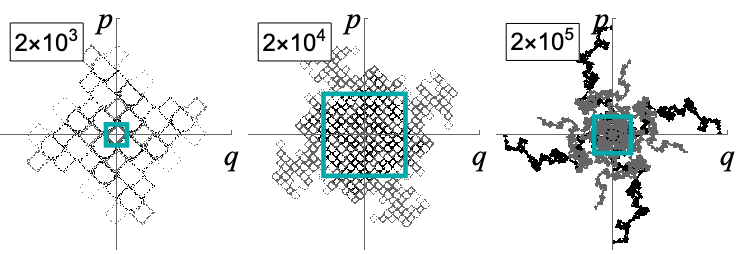}
\caption{\label{fig:Diffusion}
    Chaotic diffusion:
    each plot shows the phase space area occupied after $N$
    iterations (provided in the top-left corner) for two sample
    chaotic orbits.
    The cyan rectangle corresponds to the area occupied in the
    adjacent plot to the left, or one lattice period
    $L \times L$ for the case $N=2 \times 10^3$.
    }
\end{figure}

Beyond the larger surviving octagons that encompass both types of
tiles (marked in red), a region of chaos unfolds, where, due to the
absence of a global isolating invariant, chaotic trajectories can
diffuse between lattice periods.
For this reason, we will refer to this scenario as {\it chaotic
diffusion}.
Notably, some invariant structures still persist, such as a chain
of 9 islands comprising a line segment with a rotation number of
$2/9$ (indicated in magenta).
However, for most initial conditions, the trajectory fills the
space between the integrable fragments in an irregular manner
(depicted by black and gray dots representing two sample orbits)
and does not ``trace out'' invariant level sets, as observed in
the case of integrable diffusion.
Figure~\ref{fig:Diffusion} illustrates the trajectories of the same
two samples as shown in Fig.~\ref{fig:Example3}, but on a different
scale in both space and time.
As evident from the sequence of plots, as the number of iterations
increases, both orbits progressively diffuse out of the central cell
and expand further in the phase space.

Despite the differences between chaos in a cell and chaotic
diffusion, an important similarity exists: both spectra reveal
chaotic trajectories through the erratic behavior of the rotation
number.
In the case of chaotic diffusion, symmetry lines periodically
intersect integrable fragments alternating with chaotic regions.
This alternation is reflected in the spectra by the presence of
horizontal lines indicating global mode-locking at $1/4$ and noisy
patterns representing chaotic zones.
Refer to the two plots on the right in Fig.~\ref{fig:Example3} for
illustration.

Flowchart and step-by-step decision list for the algorithm designed
to distinguish chaotic and integrable spectra based on the assumed
piecewise-monotonicity of $\nu(s)$ are presented in
Appendix~\ref{secAPP:Algorithm}.
\vspace{-0.2cm}

\newpage

\section{\label{sec:Results} New mappings}

A systematic exploration within the same parameter space as
in \cite{ZKhNArXiV} has unveiled a notably larger number of
new integrable scenarios.
Almost all of over a hundred previously discovered systems
with regular piecewise linear forces can be in some way wrapped
on a tor.
At this point, the challenge for mappings integrable on a torus
shifts towards the classification of these new findings and
ultimately comprehending the entirety of all possibilities,
rather than merely expanding the count of discoveries.

\begin{figure}[b!]\centering
\includegraphics[width=\columnwidth]{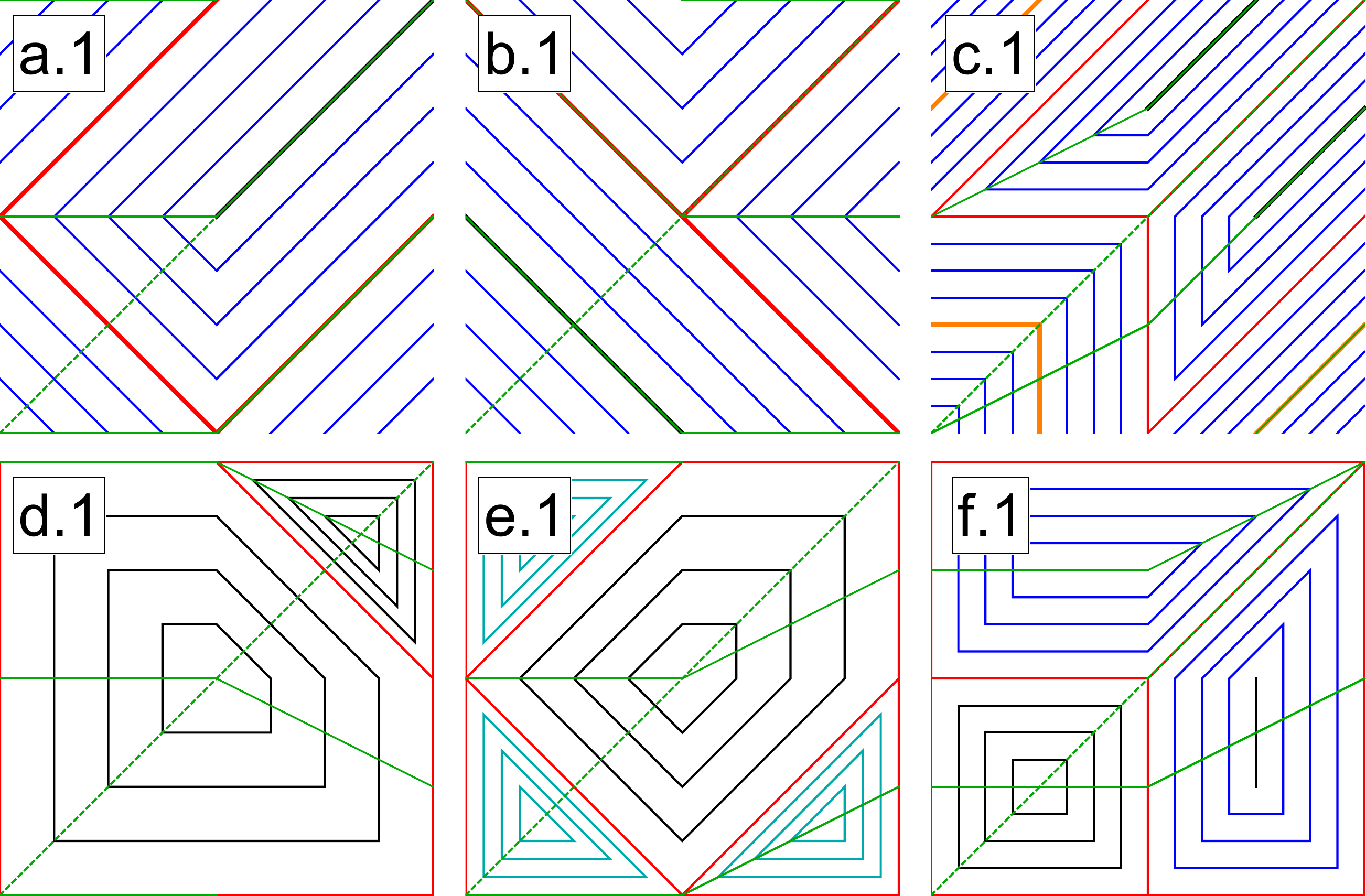}\\
\vspace{0.2cm}
\includegraphics[width=\columnwidth]{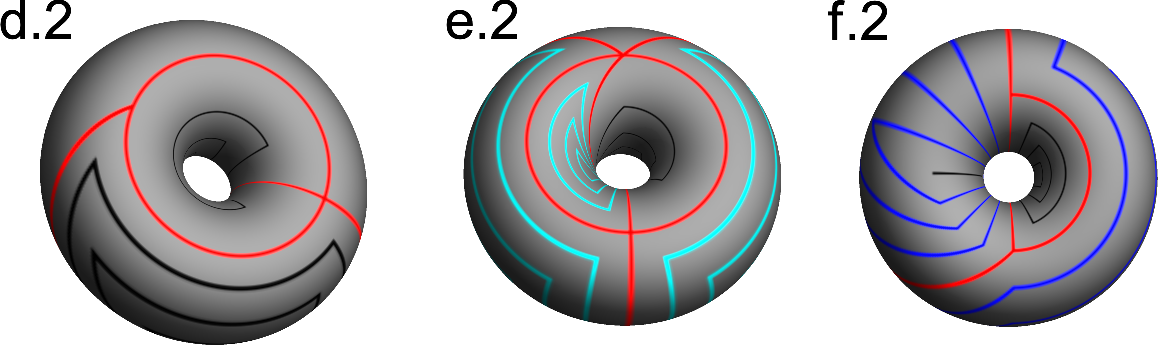}
\caption{\label{fig:Tor}
Phase space portraits for mappings integrable on a torus with
$l_{1,2}=L/2$ (a.1 – f.1) and their projections onto a toroidal
surface for corresponding cases (d.2 – f.2).
}
\end{figure}

Fig.~\ref{fig:Tor} illustrates several examples of phase
space portraits for systems integrable on a torus.
All forces are derived from the simplest piecewise linear
functions comprised of two segments of an equal length,
$L/2$.

\begin{figure}[h]\centering
\includegraphics[width=\columnwidth]{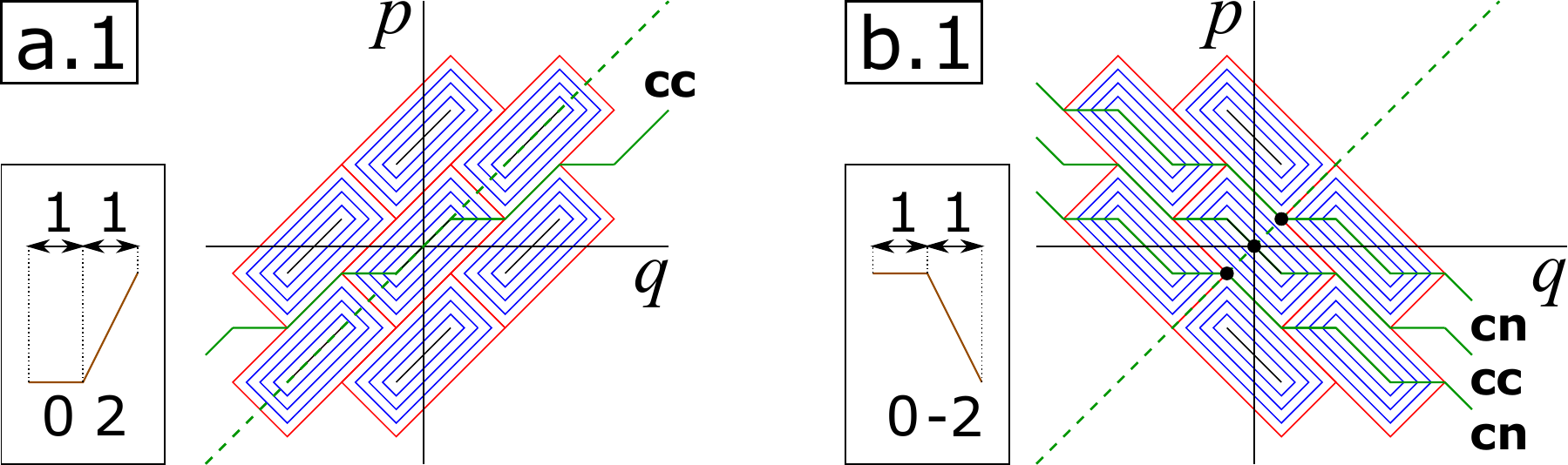}
\caption{\label{fig:AQ2piece}
    Phase space portraits for mappings $\m_\mathrm{a1}^6$ and
    $\m_\mathrm{b1}^3$.
}       
\end{figure}

\newpage
As continuous forces with arithmetic quasiperiodicity occupy a
significantly smaller set in the parameter space, we were able to
classify most new findings and present them in a compact manner.
During our parameter scan for functions composed of two pieces and
integer coefficients, no additional integrable cases were discovered,
besides the aforementioned tessellations that can be produced by
unwrapping mappings (a.1) and (b.1) from the torus, as shown in
Fig.~\ref{fig:AQ2piece}.

\begin{figure*}[p!]\centering
\includegraphics[width=0.515\linewidth]{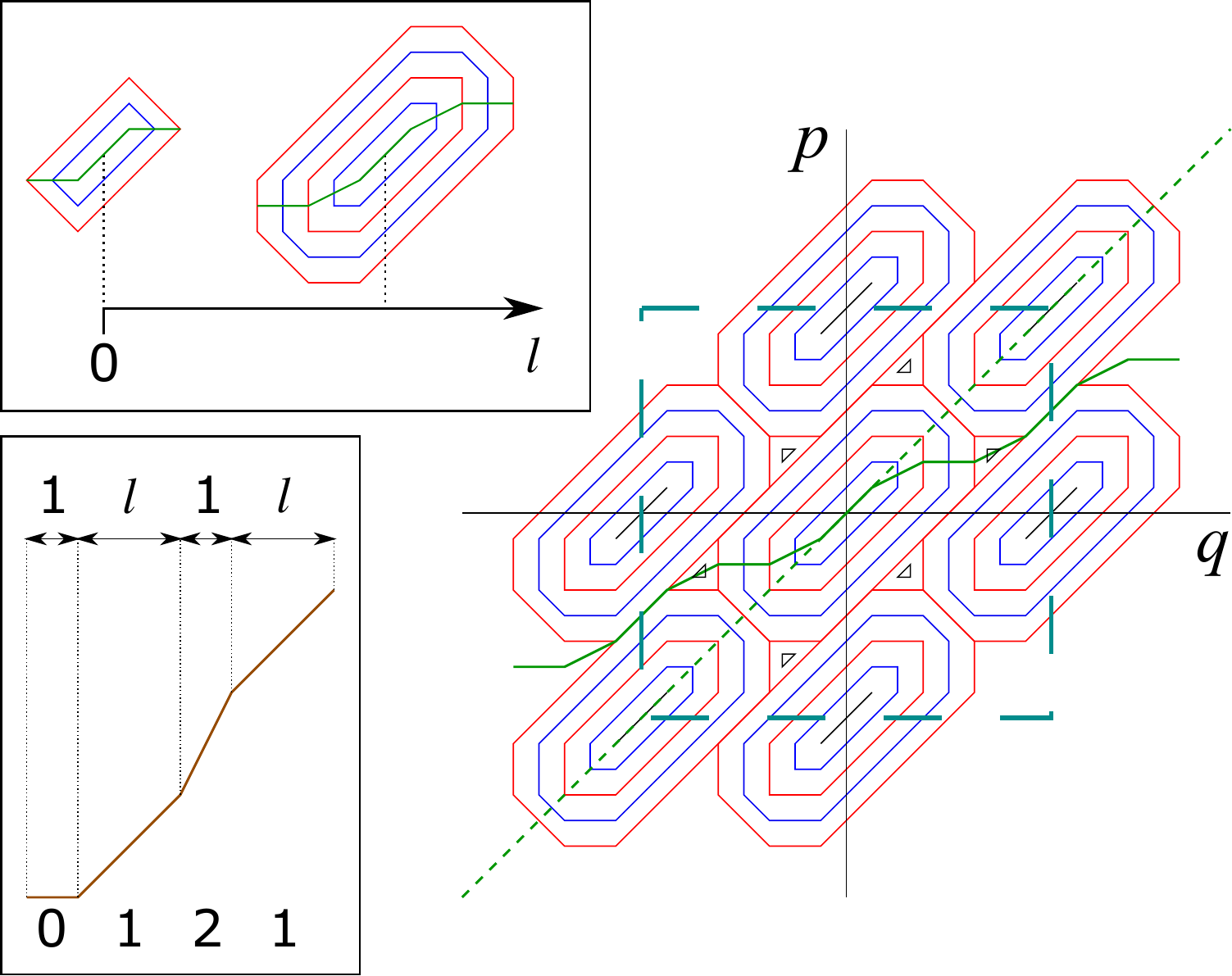}\hspace{1cm}
\includegraphics[width=0.395\linewidth]{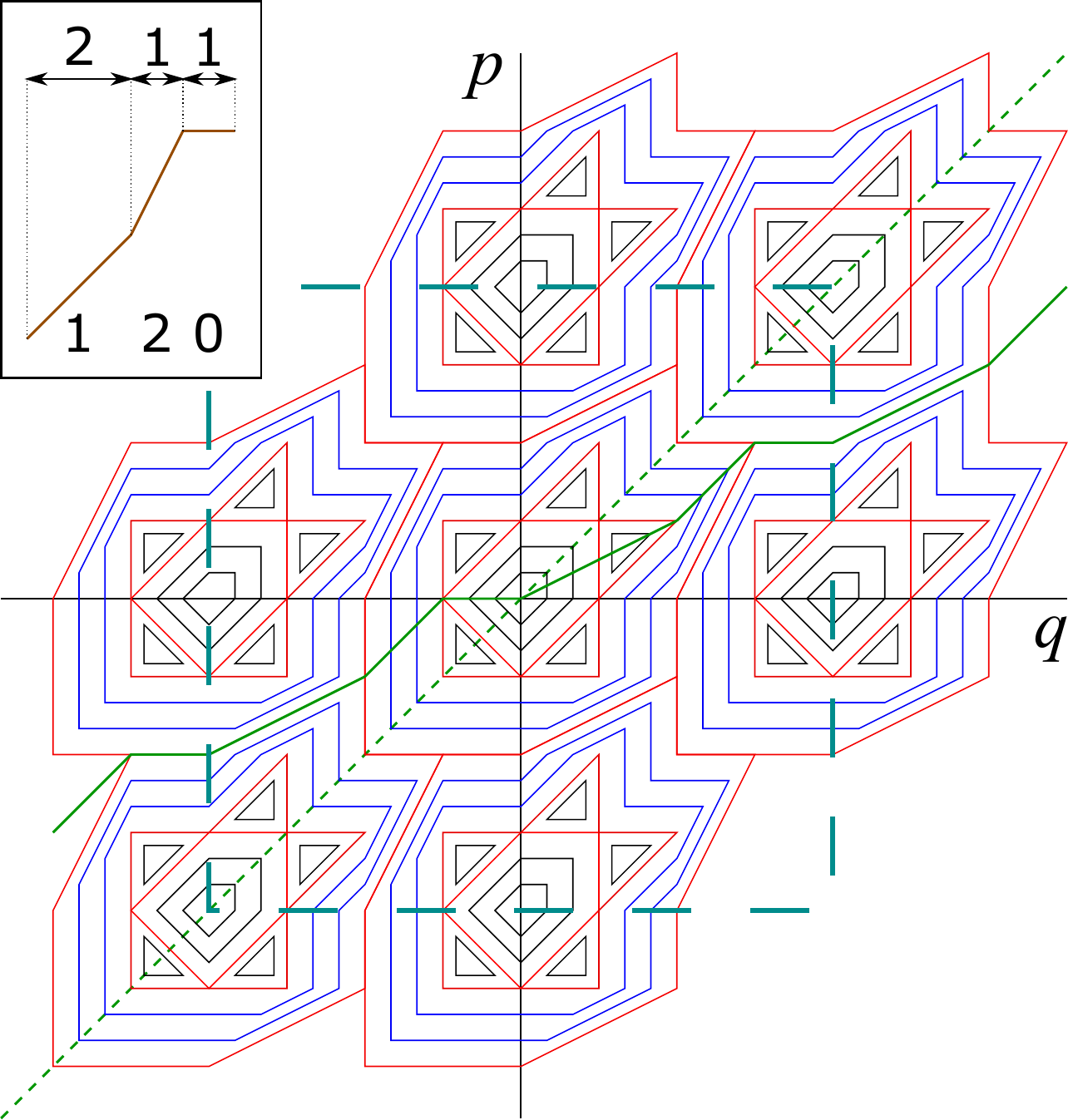}\vspace{0.2cm}
\includegraphics[width=0.975\linewidth]{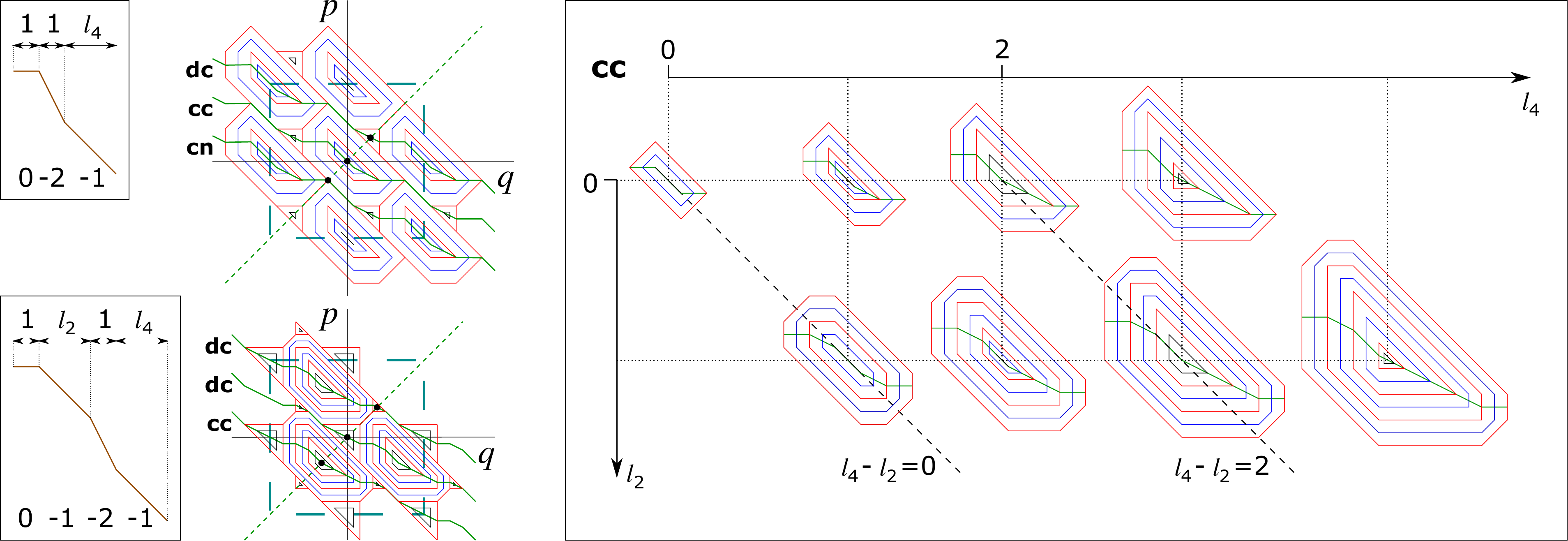}\vspace{0.2cm}
\includegraphics[width=0.975\linewidth]{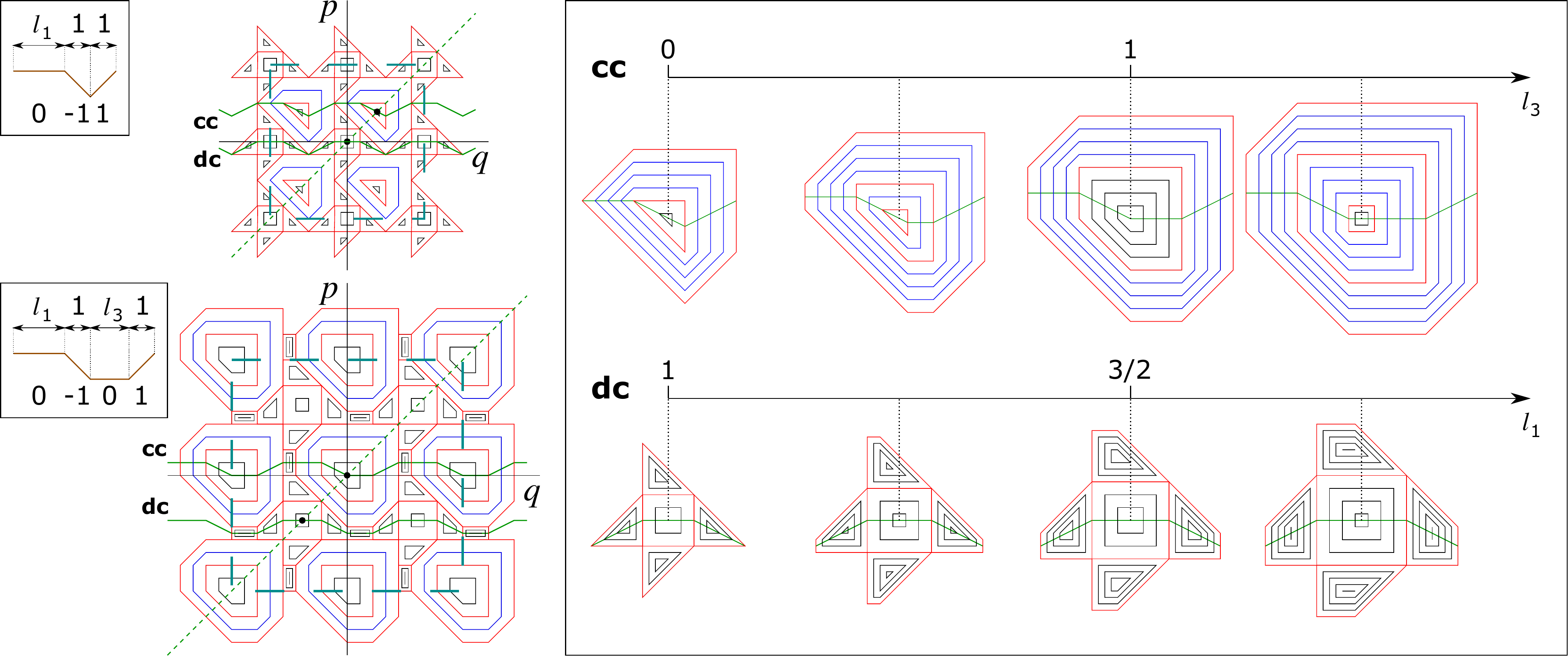}
\caption{\label{fig:MapAll}
    Phase space portraits and bifurcation diagrams for integrable
    maps with continuous arithmetically quasiperiodic forces,
    $f_\mathrm{a.q}(q)$.
    Each case is provided with one period of the force function
    (brown line) along with values of slopes and length of segments.
    Throughout this paper, we use red lines to depict separatrices,
    and blue and black to show invariant level sets within the
    areas with amplitude-dependent or linear (mode-locked) motion,
    respectively.
    Dashed and solid green lines correspond to the first and second
    symmetry lines.
}
\end{figure*}

However, as we proceeded to consider $f_\mathrm{a.q.}(q)$ formed
from three line segments, we discovered additional mappings.
Some of these systems allowed for further extension to four pieces
and/or generalization of the length of line segments to the field
of rational numbers.
Phase space portraits, along with bifurcation diagrams for the
tiles, are presented in Fig.~\ref{fig:MapAll}, while a detailed
description of the dynamics, along with analytical expressions
for rotation numbers and spectra, is provided in
Appendix~\ref{secAPP:Maps}.
Each diagram is accompanied by a sketch of one lattice period of
the force function (shown in brown), along with correponding
values of slopes and length of segments.

\section{\label{sec:Appl} Applications of mappings with periodic
and arithmetically quasiperiodic forces.}

Applications of integrable maps with polygonal invariants were
extensively discussed in~\cite{ZKhNArXiV}.
In this section, we specifically highlight some points related
to tessellations.

\subsection{\label{sbs:Suris} Conection to McMillan-Suris mappings}

The simplest 1-piece mappings on a torus with integer coefficients,
\[
    f_\mathrm{tor}(q) = k_0\,q \, \mod L,
        \qquad\qquad\quad
    k_0 = -2,-1,0,1,2,
\]
can be regarded as ultra-discrete or limiting cases of McMillan-Suris
mappings with a trigonometric polynomial
\cite{suris1989integrable,ZKhNArXiV},
\[
\begin{array}{l}
\ds \K[q,p] = \Lambda_1\left[
        \cos(p - \Phi) + \cos(q - \Phi)
    \right] +   \\[0.25cm]
\ds \qquad
    \Lambda_2 \cos(p + q - \Psi) +
    \Lambda_3 \cos(p - q)
\end{array}
\]
and force function,
\[
    f(q) = 2\,\arctan\left[
    \frac{\Lambda_1\sin\Phi + \Lambda_2 \sin(\Psi-q) + \Lambda_3 \sin q}
         {\Lambda_1\cos\Phi + \Lambda_2 \cos(\Psi-q) + \Lambda_3 \cos q}
    \right].
\]

\newpage
\begin{figure}[t!]\centering
\includegraphics[width=\columnwidth]{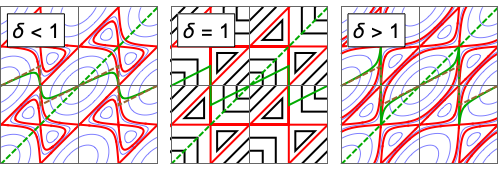}
\vspace{-0.8cm}
\caption{\label{fig:Suris}
    Phase space portraits for McMillan-Suris mappings with
    trigonometric polynomial~\ref{math:KSurisTrig}.
    Different plots correspond to different values of $\delta$.
    When $\delta \neq 1$, the dashed brown line shows linear flow
    on a torus (case $\delta = 1$) for the reference.
    }\vspace{-0.5cm}
\end{figure}

For instance, by setting $\delta = \Lambda_3/\Lambda_1$ and
$\Lambda_2 = \Phi = 0$, we obtain
\begin{equation}
\label{math:KSurisTrig}
    \K(p,q) = \cos p + \cos q + \delta\,\cos(p - q),
\end{equation}
and
\[
    f(q) = 2\,\arctan\left[
        \frac{\delta\,\sin q}{1 + \delta\,\cos q}
    \right].
\]
As seen, when $\delta = 1$, the force function reduces to
$
f(q) = 2\,\arctan\left[\tan(q/2)\right]
$.
Depending on whether we assume the continuity of $f(q)$ or take
the principal value of $\arctan$, we obtain either $f(q)=q$ or
$f(q)=q\,(\mathrm{mod}\,L)$ with $L=2\,\pi$.
This corresponds to unwrapping with arithmetic quasiperiodicity
(and zero periodic part) or to a purely periodic function with
discontinuities.
The associated invariant of motion possesses degeneracy,
resulting in infinitely many possible invariants of motion,
see~\cite{ZKhNArXiV}, including the smooth
form~(\ref{math:KSurisTrig}) and the polygon tessellation.
The middle plot in Fig.~\ref{fig:Suris} shows the polygon
invariant for $\delta=1$, surrounded by complimentary cases
with $\delta \lessgtr 1$.

\vspace{-0.3cm}
\subsection{\label{sbs:Smooth} Smoothening procedure}

\vspace{-0.2cm}
While some of these mappings offer insight into the limiting
integrable dynamics with a smooth $f(q)$, as seen in the previous
section, there is a more important application that can be used
with all discovered systems.
By ``smoothening'' $f(q)$~\cite{rychlik1998algebraic,ZKhNArXiV},
we can obtain quasi-integrable systems where the phase space
volume occupied by chaotic trajectories decreases with the
smoothness parameter, $\epsilon$.
In addition to the ability to suppress chaos, this procedure
allows the physical realization not only for forces with
arithmetic quasiperiodicity but even for periodic functions
with discontinuity, where vertical segments can be approximated
by very steep and short parts of the force function. 
In fact, Fig.~\ref{fig:Suris} provides two examples of
``perfect smoothening'' (since resulting systems are perfectly
integrable) with respect to the parameter $\epsilon = 1 - \delta$. 

\begin{figure}[h!]\centering
\includegraphics[width=\columnwidth]{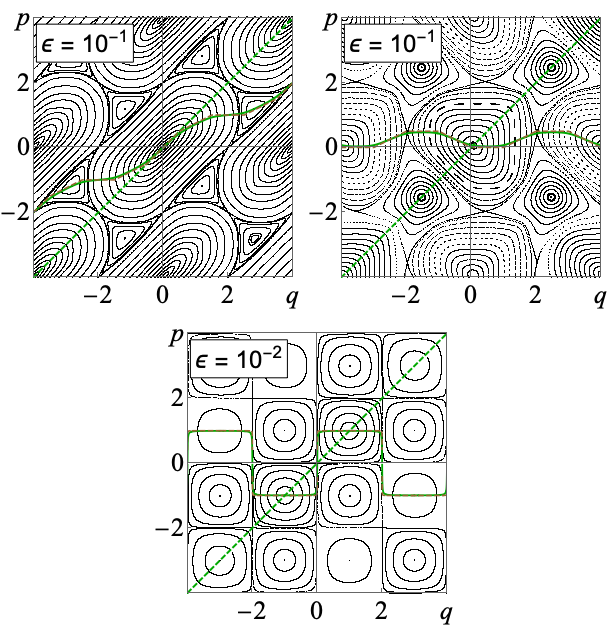}
\vspace{-0.6cm}
\caption{\label{fig:Smooth}
    Illustration of ``smoothening'' procedure.
    Examples correspond to forces with arithmetic quasiperiodicity
    $\m^6_\mathrm{a2}$ (top left), purely periodic
    $\m^4_\mathrm{F1'}$ (top right), and periodic function with 
    discontinuity, square wave (plot at the bottom).
    }
\end{figure}

By using other rules, as long as the new function is close enough
to its piecewise linear prototype, we can obtain near-integrable
dynamics.
Using 4-piece mappings $\m^6_\mathrm{a2}$ and $\m^4_\mathrm{F1'}$
for illustration, if all segments have equal unity length, we can
approximate forces as
\[
f^\mathrm{a2}_\mathrm{a.q}(q) = \Delta[q+1]+q
\qquad\text{and}\qquad
f^\mathrm{F1'}_\mathrm{a.q}(q) = \Delta[q],
\]

\noindent
where we introduced smooth differentiable form
(when $\epsilon \neq 0$) of the trapezoid wave:
\[
\Delta[q] = 
\frac{
    \arccos\left[(1-\epsilon)\,\cos\frac{\pi\,q}{2}\right] -
    \arcsin\left[(1-\epsilon)\,\sin\frac{\pi\,q}{2}\right]
}{\pi}.
\]
Figure~\ref{fig:Smooth} provides an illustration of phase space
portraits for these systems obtained by tracking for
$\epsilon = 10^{-1}$.
The last example at the bottom depicts tracking for the smooth
square wave force
\[
\Pi[q] = \frac{4}{\pi}\,
    \arctan\left[\frac{1}{\epsilon}\,\sin\frac{\pi\,q}{2}\right]
\]
which, for $\epsilon = 0$, represents another limiting case of the
McMillan-Suris map with $k_0 = 0$.

In particular, this method allows the design of nonlinear
quasi-integrable kick functions corresponding to thin RF cavities
when considering a simplified picture of longitudinal motion in
particle accelerators.
Even for a single degree of freedom, as highlighted by Chirikov's
early findings~\cite{chirikov1969research,chirikov1979universal},
the typical kick from the RF field on a particle
($\propto\sin q$, taking $q$ as the longitudinal coordinate)
results in dynamics that exhibit chaos and can lead to global
stochasticity.
The smoothening procedure, indeed, provides an alternative to the
only known integrable case of the McMillan-Suris map.
Cases with $F=0$ represent longitudinal dynamics for the stationary
bucket (indicating no acceleration), while $F \neq 0$ represents
regimes with net acceleration/deceleration.

\newpage
\section{\label{sec:Summary} Summary}

In this paper, we have uncovered a novel class of integrable
symplectic maps characterized by a polygonal tessellation and
corresponding fibration of the phase space.
While polygonal tilings of the plane have found applications in
diverse fields such as crystallography, the study of topological
phases of matter, and group theory, our work introduces a distinct 
category of periodic regular and semi-regular tessellations
directly linked to the presence of integrability.
Our investigation focuses on systems in the McMillan-H\'enon form,
featuring two specific types of piecewise linear force functions
with patterns of regularity, namely, (i) functions with arithmetic
quasiperiodicity and (ii) periodic functions with discontinuities.
The mentioned subsets collectively form a broader category of
systems defined on a torus.

These nonlinear maps manifest non-trivial dynamical regimes,
encompassing various configurations of the center, mode-locking,
and integrable diffusion.
In the latter scenario, which has not been previously documented
to the best of our knowledge, a particle quasi-randomly hops
between tiles, gradually ``diffusing'' away while remaining
constrained to a constant level set of invariant.

To conduct a systematic search for such maps, we developed an
automated algorithm that evaluates the rotation number along
symmetry lines.
Integrable systems of this kind are distinguished by
piecewise monotonic behavior of rotation number concerning
initial amplitude.

\newpage
\noindent
In contrast, chaotic systems display irregular and ``jittery''
behavior, enabling a clear distinction between the two.
Our search focused on force functions with integer slope
coefficients, however, it remains an open question whether
integrable maps with polygonal invariants also exist for
rational, and especially irrational, slope coefficients.
While we have presented numerous new mappings throughout this
paper, our algorithm, especially when not restricted to arithmetic
quasiperiodicity, unveils many more systems than we can present
in a methodical and coherent manner.
This emphasizes the importance of a more fundamental challenge:
providing all possible integrable configurations from the first
principles.
Another interesting future direction would be to study a quantum
version of the discovered integrable maps.

\section{Acknowledgments}

The authors would like to thank Taylor Nchako (Northwestern
University) for carefully reading this manuscript and for her
helpful comments.
This manuscript has been authored by Fermi Research Alliance, LLC
under Contract No. DE-AC02-07CH11359 with the U.S. Department of
Energy, Office of Science, Office of High Energy Physics.
Work supported by the U.S. Department of Energy, Office of Science, Office of Nuclear Physics under contract DE-AC05-06OR23177.

\newpage
$\phantom{d}$

\newpage
$\phantom{d}$


%

\appendix

\newpage
\begin{figure}[t!]\centering
\includegraphics[width=\columnwidth]{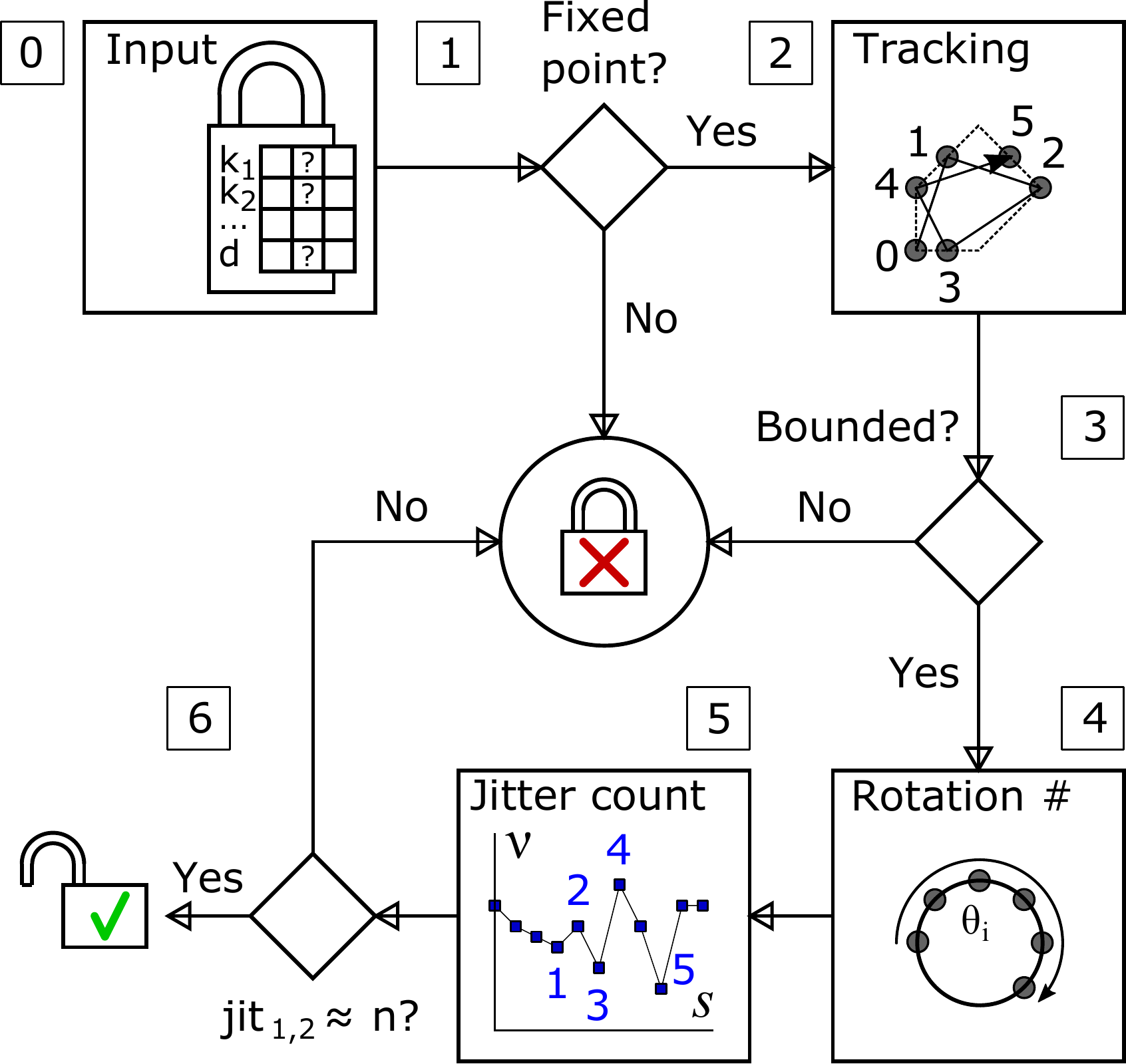}
\caption{\label{fig:PRF_Algorithm}
    Flowchart representing the machine-assisted discovery of
    integrable maps with polygonal tessellation.
    If the map defined by the input (Step 0) has a fixed point
    (Step~1), the algorithm performs tracking for different initial
    conditions (Step 2).
    Then, if all orbits are bounded (Step 3), it proceeds to the
    evaluation of the rotation number along both symmetry lines
    (Step 4), and then counts the number of times $\nu$ changes
    monotonicity (Step 5).
    Finally, if $\mathrm{jit}$ is comparable to the number of
    segments in the force function (Step 6), the map is sent
    for analytic verification.
    The search process terminates when any conditional operation
    is false (NO).
    }
\end{figure}

\section{\label{secAPP:Algorithm} Search Algorithm}

To measure the monotonicity of the rotation number, we introduce
a variable, $\mathrm{jit}$, which reveals rapid oscillations or
deviations in the rotation number from the anticipated piecewise
monotonic pattern characterizing an integrable scenario.
In the context of a completely integrable system, the value of
$\mathrm{jit}$ is smaller than or equal to the number of layers
with different types of polygons, which is approximately
equivalent to the number of segments in $f(q)$:
\[
    \mathrm{jit} \leq \text{number of layers} \approx n.
\]
In contrast, for scenarios involving chaotic behavior,
$\mathrm{jit}$ is roughly analogous to the number of chaotic
samples within the layer
\[
    \mathrm{jit} \approx \text{number of chaotic samples}.
\]
By densely populating each segment along symmetry lines with
sample points, we can effectively distinguish between these
two behaviors.
The corresponding algorithm is visualized in the flowchart
presented in Fig.~\ref{fig:PRF_Algorithm}, while specific
processes and decision points are outlined below.

\newpage
\begin{enumerate}
\item[0.]{\bf Input.}
    The search process involves exploring different force
    functions $f(q,\mathbf{k},\mathbf{l},d)$ with integer vectors
    $\mathbf{k}$, $\mathbf{l}$, and scalar $d$.
    While the initial scan focused on integer slopes $\mathbf{k}$,
    some results can be generalized analytically for more general
    values of the shift parameter $d\in\mathbb{R}$ and segment
    lengths $\mathbf{l}\in\mathbb{R}_+^n$.
\item[1.]{\bf Fixed point.}
    A crucial requirement for organizing the phase space into cells
    of periodicity is the existence of at least one fixed point
    $\zeta_* = (q_*,p_*)$, where $q_* = p_* = f(q_*)/2$.
    Maps lacking a fixed point are excluded from consideration.
\item[2.]{\bf Orbit tracking.}
    For the chosen maps, we trace their orbits
    $\vec{\zeta} = (\zeta_0,\ldots,\zeta_N)$ for various initial
    conditions $\zeta_0 = (q_0,p_0)$ along both symmetry lines:
    \[
        p_0 = q_0
        \qquad\quad\mathrm{and}\qquad\quad
        p_0 = f(q_0)/2.
    \]
    Initial points are placed within two lattice periods around
    the fixed point, $|q_0-q_*|<L$, ensuring multiple seeds for
    every $j$-th segment, i.e.,
    $q_0\in(\sum_{i=1}^j l_i;\sum_{i=1}^{j+1} l_i)$.
\item[3.]{\bf Stability check.}
    Although mappings possess fixed points, they might still
    exhibit unstable trajectories.
    If we observe unbounded growth of the radial distance from
    the fixed point for certain initial conditions, i.e.,
    $\exists\,i:\,\,|\zeta_i - \zeta_*| > r_\mathrm{max}$, the
    map is excluded from further analysis.
    In particular, mappings with chaotic diffusion might be
    rejected at this stage.
    We set $\ds r_\mathrm{max}=4\,L$.
\item[4.]{\bf Rotation number calculation.}
    Next, we evaluate the rotation number for each initial
    condition as:
    \[
        \nu \approx \frac{\theta_N - \theta_0}{2\,\pi\,N}
    \]
    where sufficiently large $N$ ensures convergence for
    invariant level sets.
\item[5.]{\bf Jitter count.}
    To quantify the variable $\mathrm{jit}$, we count the number
    of times the rotation number changes its monotonicity as a
    function of the coordinate, and calculated along both symmetry
    lines
    \[
    \begin{array}{r}
    \ds\mathrm{jit}_{1,2} = \frac{1}{2}\,\sum_{|q_i-q_*| \leq L}
    \qquad\qquad\qquad\qquad\qquad\qquad\qquad    \\
    \ds \qquad\qquad \Big[ 1 - \mathrm{sgn}\,\big\{
            [\nu(q_i)-\nu(q_{i-1})]
            [\nu(q_{i+1})-\nu(q_i)]
        \big\} \Big].
    \end{array}
    \]
\item[6.]{\bf Selection}
    In the final stage, we compare the value of $\mathrm{jit} =
    \mathrm{max}\left[ \mathrm{jit}_1, \mathrm{jit}_2 \right],$
    with the number of samples in all layers.
    If
    \[
    \mathrm{jit} \approx n
    \ll \text{number of samples in layers}
    \] 
    we select the map and then analytically verify its
    integrability.
\end{enumerate}

\section{\label{secAPP:Maps}
Integrable mappings allowing \\
forces with arithmetic quasiperiodicity}

\begin{figure}[t!]\centering
\includegraphics[width=\columnwidth]{PRF_Map1516.pdf}\\
\vspace{0.2cm}
\includegraphics[width=\columnwidth]{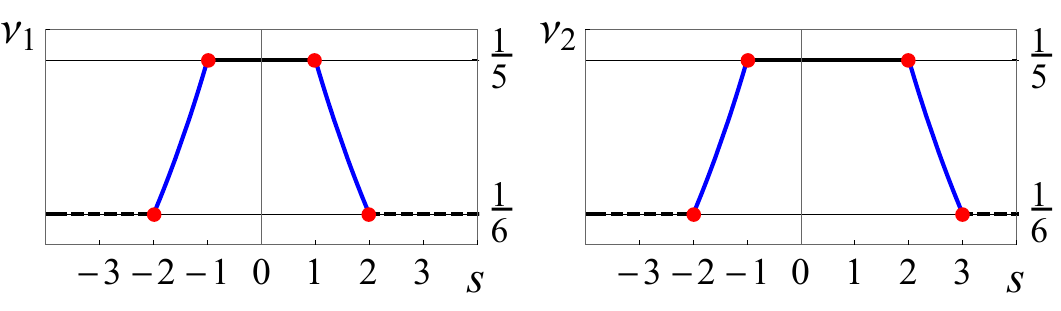}
\vspace{-0.5cm}
\caption{\label{fig:Map1516}
    Phase space portrait (top) and the corresponding spectra along
    the first ($\nu_1$) and second ($\nu_2$) symmetry lines for the
    map $\m_\mathrm{G}^6$.
}
\end{figure}

\subsection{Map $\m_\mathrm{G}^6$}

Proceeding to force functions formed from three line segments, we
encountered an intriguing map
\[
\m_\mathrm{G}^6:\qquad
    f(q) = [\{1;2\},\{2;1\},\{0;1\};4\,m].
\]
This system doesn't allow for any choice in selecting the
configuration of the center and doesn't contain any free
parameters except for scaling and trivial translations.
The phase space diagram and rotation number along symmetry lines
are depicted in Figure~\ref{fig:Map1516}, where $\nu_{1,2}$ are
given by:
\[
\nu_1 =
\left\{
\begin{array}{lr}
\ds \frac{1}{5},                &\,\,
\ds |s| \leq 1,                 \\[0.35cm]
\ds \frac{1}{5+(|s|-1)},        &\,\,
\ds 1 < |s| \leq 2,
\end{array}
\right.
\]
and
\[
\nu_2 =
\left\{
\begin{array}{lr}
\ds \frac{1}{5},                &\,\,
\ds |s-1/2| \leq 3/2,           \\[0.35cm]
\ds \frac{1}{5+(|s-1/2|-3/2)},  &\,\,
\ds 3/2 < |s-1/2| \leq 5/2.
\end{array}
\right.
\]

The central cell has two layers:
the inner core consists of nested pentagons surrounded by a chain
of 5 degenerate triangular islands (level sets shown in black).
All initial conditions falling within this layer exhibit periodic
behavior with a rotation number equal to $1/5$.
On the other hand, the outer layer is composed of concentric
dodecagons (level sets shown in blue) and exhibits nonlinear
integrable motion, which includes periodic or quasiperiodic
orbits.
The surrounding cells are organized into groups of 6, thus
characterizing global mode-locking at a rotation number of $1/6$.

\subsection{Mappings
    $\m_{\alpha 3}^4$ and $\m_{\mathrm{F}1'}^4$}

The next integrable system we discovered for forces composed of
three segments is the mapping $\m_{\alpha 3}^4$:
\[
\begin{array}{ll}
\mathrm{[cc]}:
& [\{0;l_1\},\{-1;1\},\{1;1\};2\,m\,(2+l_1)-2\,\,],
\\[0.25cm]
\mathrm{[dc]}:
& [\{0;l_1\},\{-1;1\},\{1;1\};2\,m\,(2+l_1)+l_1],
\end{array}
\]
with two possible configurations of the central cell and one real
parameter, $l_1 \geq 1$.
By adding a fourth piece, it can be generalized to another map,
$\m_{\mathrm{F}1'}^4$:
\[
\begin{array}{rl}
\mathrm{[cc]}:
& [\{0;l_1\},\{-1;1\},\{0;l_3\},\{1;1\};
        d_* - 2 - l_3],
\\[0.25cm]
\mathrm{[dc]}:
& [\{0;l_1\},\{-1;1\},\{0;l_3\},\{1;1\};
       d_* + l_1],
\end{array}
\]
with two parameters $l_1 \geq 1$, $l_3 > 0$, and where
\[
d_* =  2\,m\,(2+l_1+l_3).
\]

\begin{figure}[t!]\centering
\includegraphics[width=\columnwidth]{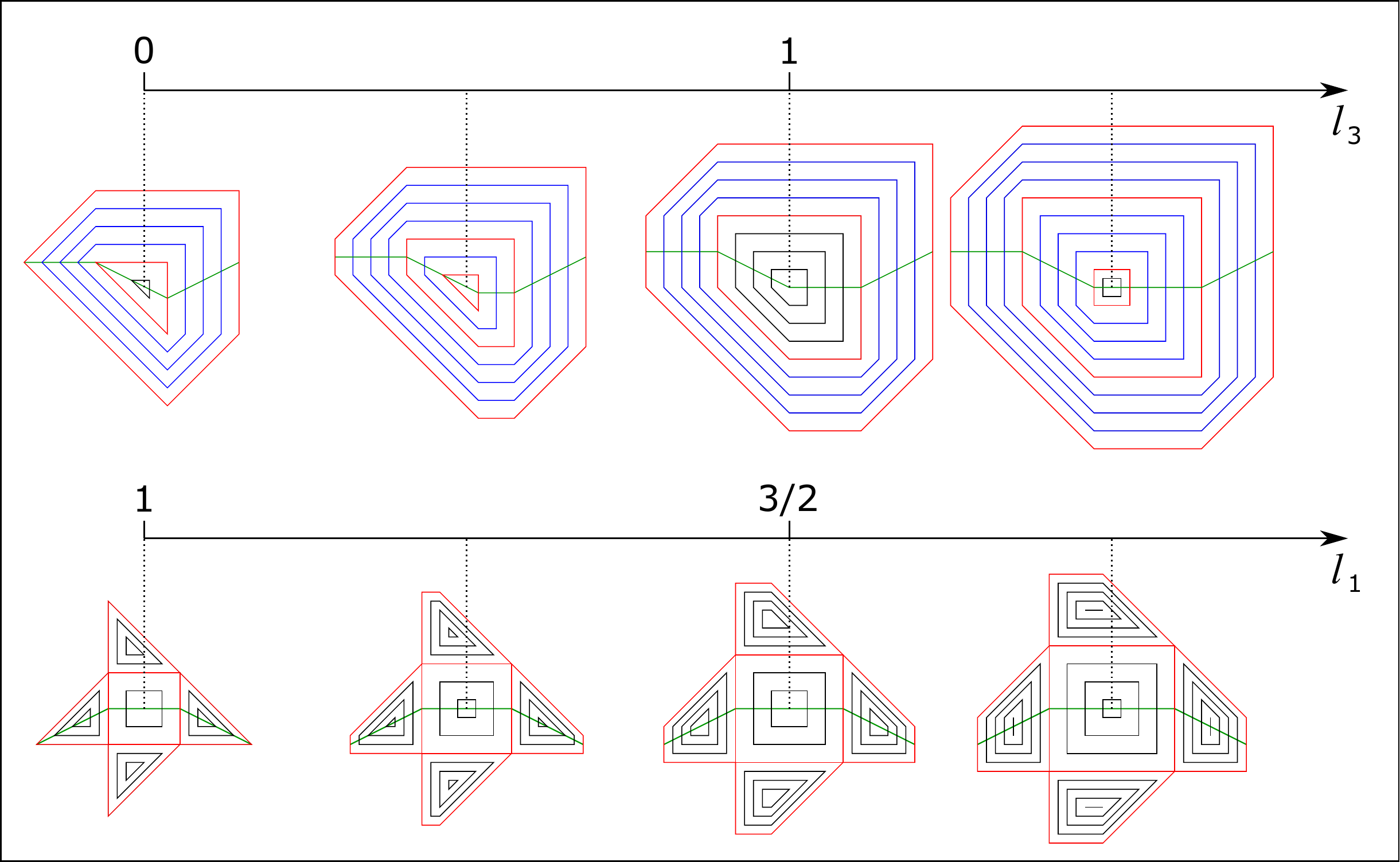}\\
\vspace{0.3cm}
\includegraphics[width=\columnwidth]{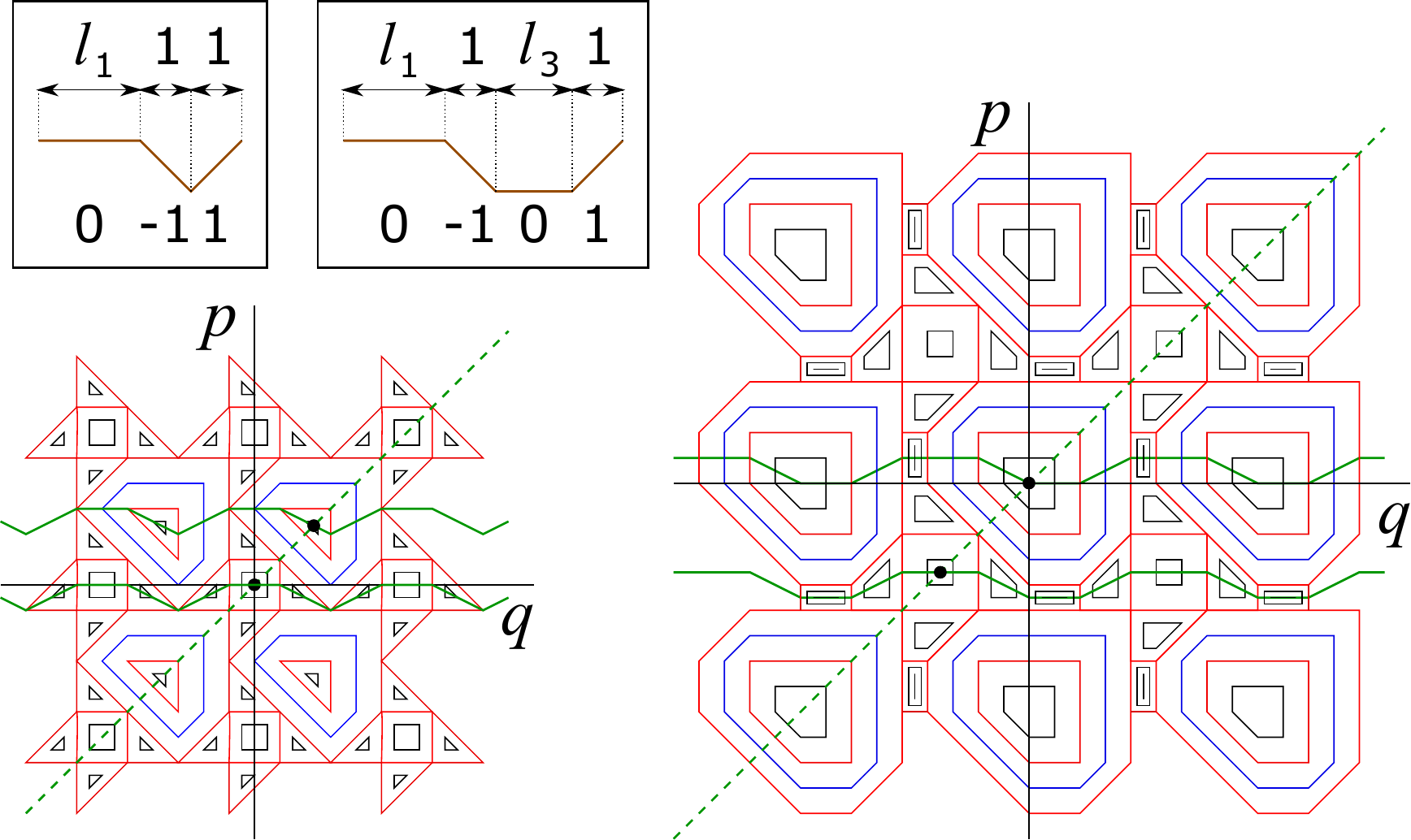}\\
\vspace{0.3cm}
\includegraphics[width=\columnwidth]{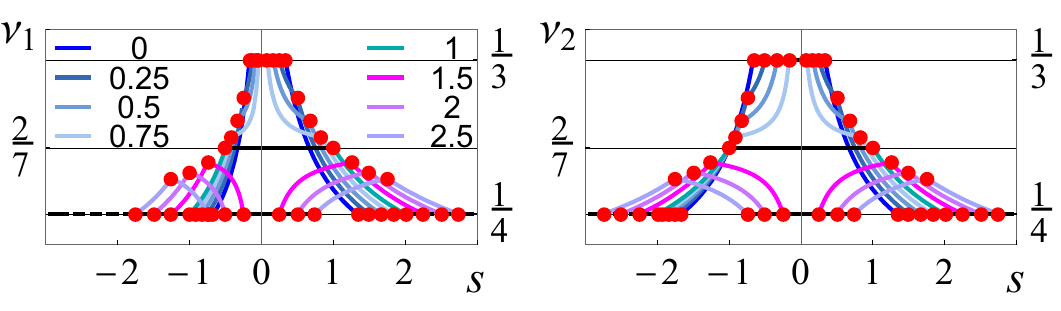}
\vspace{-0.5cm}
\caption{\label{fig:Map1414}
    Bifurcation diagram (top) and phase space portraits (middle)
    for mappings $\m_{\alpha 3}^4$ and $\m_{\mathrm{F}1'}^4$.
    The bottom row shows spectra for the cc configuration
    and different values of $l_3$ (shown in different colors).
    }\vspace{-0.8cm}
\end{figure}

Both systems have $k_0=0$, resulting in a global period and
rotation number equal to $4$ and $1/4$, respectively.
The tiling is semi-regular with two types of cells grouped in
chains of four islands around the central cell
(see Fig.~\ref{fig:CCCN} for the reference).
The configuration labeled as {\it dc} stands for {\it degenerate
center} and can be seen either as a central cell where all initial
conditions are mode-locked at a global rotation number, or as a
``thick'' central node that is part of a net of separatrices.
Phase space plots, bifurcation diagrams for tiles, and examples
of the rotation number for different values of parameters are
provided in Fig.~\ref{fig:Map1414}.
Note that changes in the parameter $l_1$ only affect the
degenerate cell, while changes in $l_3$ only impact the cell with
nonlinear dynamics.
For analytical expressions of rotation numbers, the reader can
refer to mappings $\alpha 3$ and $\mathrm{F}1'$
in~\cite{ZKhNArXiV}.

\subsection{Mappings
    $\m_{\mathrm{E} 1'}^3$, $\m_{\mathrm{a} 2}^6$
    and $\m_{\mathrm{b} 2}^3$}

\begin{figure}[t!]\centering
\includegraphics[width=\columnwidth]{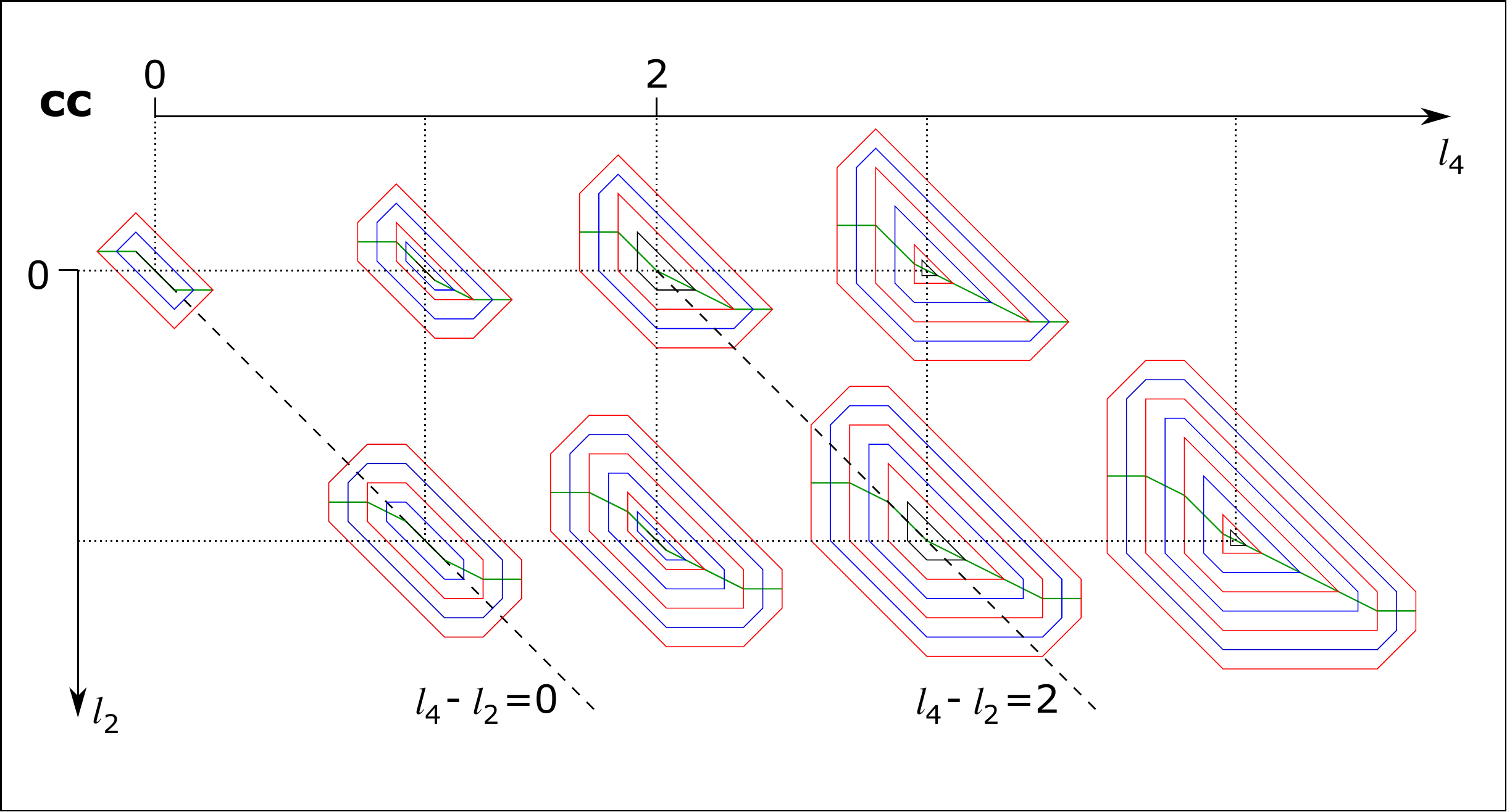}\\
\vspace{0.3cm}
\includegraphics[width=\columnwidth]{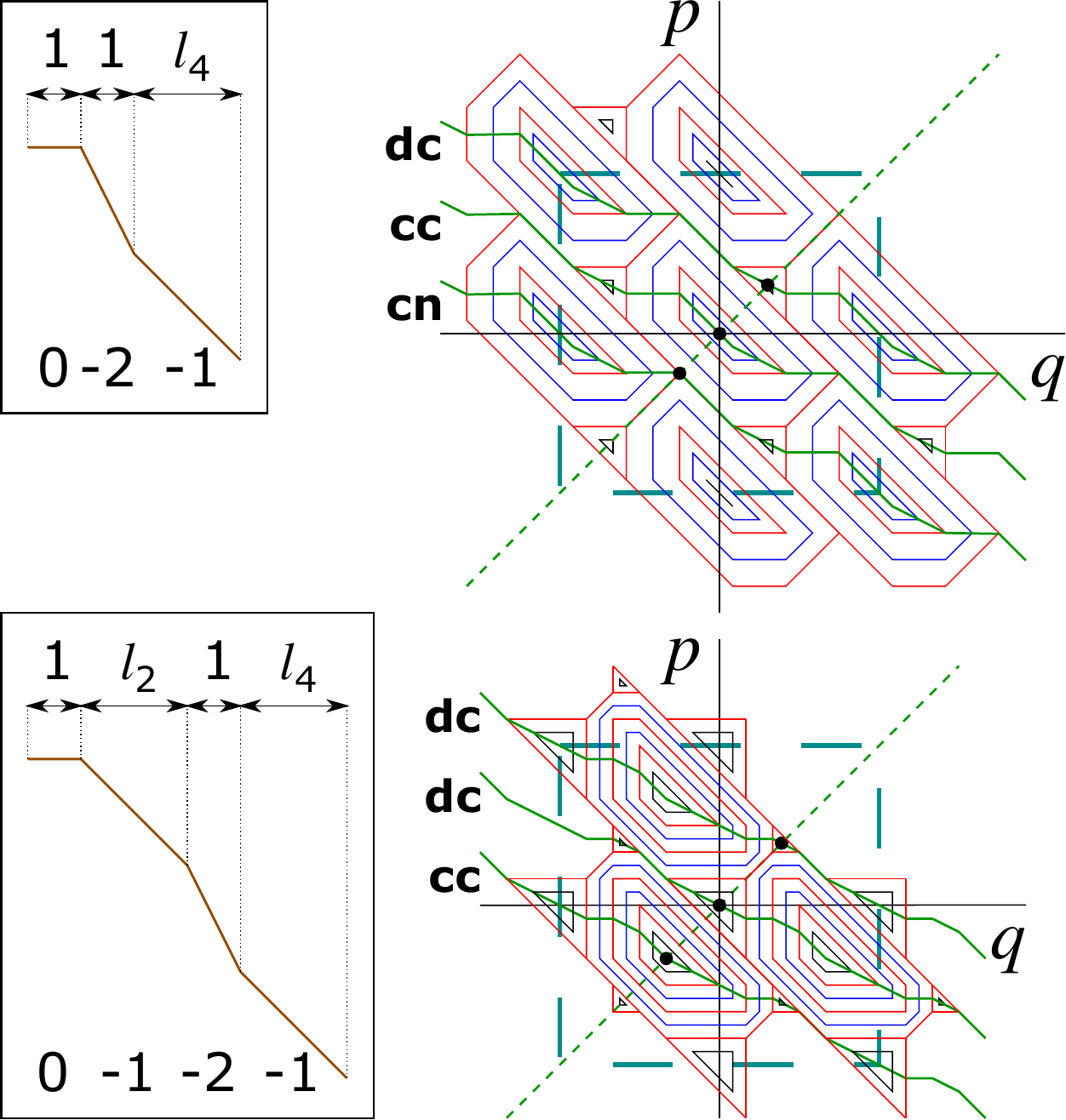}
\caption{\label{fig:Map1213}
	Phase space portraits and bifurcation diagram for mappings
        $\m_{\mathrm{E} 1'}^3$ and
        $\m_{\mathrm{b} 2 }^3$.
        }
\end{figure}

The final group of systems to be presented in this section
involves the extension of the mapping
$\m_{\mathrm{a} 1}^6$ to a force function composed
of four pieces and $\m_{\mathrm{b} 1}^3$ to forces
made of three or four segments.
The first extension is $\m_{\mathrm{a} 2}^6$:
\[
\mathrm{[cc]}:\,
[\{0,1\},\{1,l\},\{2,1\},\{1,l\};2\,m\,(1+l)-l],
\quad l > 0,
\]
where the new segments should have equal length, i.e.,
$l_2 = l_4 = l$. 
Two other mappings are $\m_{\mathrm{E} 1'}^3$
\[
\begin{array}{ll}
\mathrm{[cc]}:              &
[\{0,1\},\{-2,1\},\{-1,l_4\};
    3\,m \,(2 + l_4) + 4 + l_4],    \\[0.25cm]
\mathrm{[cn]}:              &
[\{0,1\},\{-2,1\},\{-1,l_4\};
    3\,m \,(2 + l_4) + 2],          \\[0.25cm]
\mathrm{[dc]}:              &
[\{0,1\},\{-2,1\},\{-1,l_4\};
    3\,m \,(2 + l_4) - l_4],
    \end{array}
\]
and $\m_{\mathrm{b} 2}^3$
\[
\begin{array}{ll}
\mathrm{[cc]}:  &\!\!\!
[\{0,1\},\{-1,l_2\},\{-2,1\},\{-1,l_4\};
    d_* + 4 + 2\,l_2 + l_4],                    \\[0.25cm]
\mathrm{[dc]}:  &\!\!\!
[\{0,1\},\{-1,l_2\},\{-2,1\},\{-1,l_4\};
    d_* + 2 + l_2],                             \\[0.25cm]
\mathrm{[dc]}:  &\!\!\!
[\{0,1\},\{-1,l_2\},\{-2,1\},\{-1,l_4\};
    d_* - l_4],
\end{array}
\]
defined for $l_4 \geq l_2 > 0$ and \[d_* = 3\,m \,(2 + l_2 + l_4).\]
The $l_4 \geq l_2$ condition  ensures the exclusion of twin maps.

\begin{figure}[t!]\centering
\includegraphics[width=\columnwidth]{PRF_Map0116.pdf}\\
\vspace{0.1cm}
\includegraphics[width=\columnwidth]{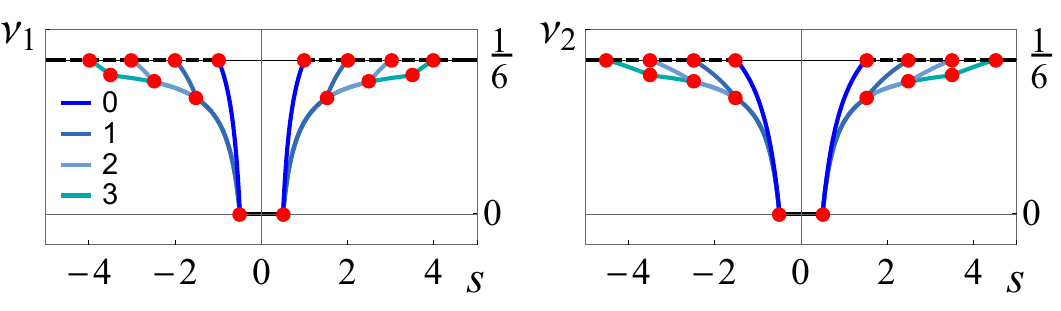}
\vspace{-0.5cm}
\caption{\label{fig:Map0116}
    Phase space portrait for the map
    $\m_{\mathrm{a} 2}^6$.
    Bifurcation diagram and spectra for different values of
    parameter $l$ (shown with different colors) are provided
    to the left and right respectively.
    }
\end{figure}

\vspace{0.28cm}
Figures~\ref{fig:Map1213} and \ref{fig:Map0116} provide
bifurcation diagrams for cells with nonlinear dynamics along with
phase space plots.
In all systems, the tiling is semi-regular: the plane is
tessellated with octagon or hexagon separatrices (red contours),
enclosing cells with nonlinear dynamics.
The gaps in between are filled with triangles representing the
degenerate center and/or chains of linear islands
(black contours).

\newpage

The map $\m_{\mathrm{a} 2}^6$ has a singular
configuration, featuring a central cell surrounded by a chain of
six identical islands.
On the other hand, $\m_{\mathrm{b} 2}^3$ allows for three
distinct configurations and encompasses two independent groups of
chains, each composed of three islands.
In the case of the latter, the fixed point can be situated either
within the central cell or within one of the two triangular tiles
representing degenerate centers.
When $l_2 = 0$ (map $\m_{\mathrm{E} 1'}^3$), one of the
triangular chains vanishes, leading to the corresponding degenerate
center evolving into a central node.
Moreover, the bottom plots in Figure~\ref{fig:Map0116}
showcases sample spectra for various parameter values of $l$.
The expressions for $\nu_{1,2}$ are as follows:
\[
\nu_j =
\left\{
\begin{array}{ll}
\ds 0,                                                      &\,\,
\ds 0 \geq |s| - \frac{1}{2},                               \\[0.35cm]
\ds \frac{\tau}{2+6\,\tau},                                 &\,\,
\ds 0 < |s| - \frac{1}{2} \leq  l,                          \\[0.35cm]
\ds \frac{j\,l+2\,(\tau-l)}{j\,(2+6\,l)+8\,(\tau-l)},       &\,\,
\ds  l < |s| - \frac{1}{2} \leq \frac{j}{2} +l,
\end{array}
\right.
\]
where $\tau = |s| - 1/2$ and $j=1,2$.

\end{document}